%% file: main.tex
\documentclass[pdflatex,sn-nature,Numbered]{sn-jnl}

\usepackage{algorithm}
\usepackage{algorithmicx}
\usepackage{algpseudocode}
\usepackage{amsmath,amssymb,amsfonts}
\usepackage{amsthm}
\usepackage[title]{appendix}
\usepackage{booktabs}
\usepackage{graphicx}
\usepackage{listings}
\usepackage{longtable}
\usepackage{manyfoot}
\usepackage{mathrsfs}
\usepackage{multirow}
\usepackage{tablefootnote}
\usepackage{textcomp}
\usepackage[table,x11names,dvipsnames]{xcolor}
\usepackage{xspace}
\usepackage{caption}
\usepackage{ragged2e}

\usepackage{CJK}
\usepackage[whole]{bxcjkjatype} 
\usepackage{CJKutf8}

\definecolor{mygray}{gray}{0.9}

\graphicspath{{figures/}}

\unnumbered

\definecolor{DarkOrange}{RGB}{204, 85, 0}
\definecolor{LincolnGreen}{RGB}{17, 102, 0}
\definecolor{Rust}{HTML}{9B4F0F}
\definecolor{DarkCyan}{HTML}{008B8B}
\definecolor{MediumAquaMarine}{HTML}{66CDAA}
\definecolor{Maroon}{HTML}{800000}
\definecolor{Crimson}{HTML}{DC143C}

\newcommand{\galex}{\textit{GALEX}}
\newcommand{\swift}{\textit{Swift}\xspace}

\newcommand{\ion}[2]{#1\,\textsc{#2}}
\newcommand{\nodata}{...}
\newcommand{\program}[1]{\texttt{#1}}

\newcommand{\at}{2021yfj\xspace}
\newcommand{\sn}{SN\,2021yfj\xspace}

\newcommand{\method}{\textit{Methods Section}\xspace}
\newcommand{\methods}{\textit{Methods Sections}\xspace}
\newcommand{\extmat} {\textit{Extended Material}\xspace}
\newcommand{\extmattwo} {Extended Material\xspace}
\newcommand{\suppmat}{\textit{Supplementary Material}\xspace}
\newcommand{\suppmattwo}{Supplementary Material\xspace}

\newcommand{\figuretwo}{\textit{Figure}\xspace}
\newcommand{\figuretwos}{\textit{Figures}\xspace}
\newcommand{\tabletwo}{\textit{Table}\xspace}

\newcommand{\araa}{Annu. Rev. Astron. Astrophys.} 
\newcommand{\aj}{Astron. J.} 
\newcommand{\apj}{Astrophys. J.} 
\newcommand{\apjl}{Astrophys. J. Lett.} 
\newcommand{\apjs}{Astrophys. J. Suppl. Ser.} 
\newcommand{\aap}{Astron. Astrophys.} 
\newcommand{\aapr}{Astron. Astrophys. Rev.} 
\newcommand{\mnras}{Mon. Not. R. Astron. Soc.} 
\newcommand{\nat}{Nature} 
\newcommand{\prl}{Phys. Rev. Lett.} 
\newcommand{\pasa}{Publ. Astron. Soc. Aust.} 
\newcommand{\pasj}{Publ. Astron. Soc. Jpn} 
\newcommand{\pasp}{Publ. Astron. Soc. Pac.} 
\newcommand{\ssr}{Space Sci. Rev.} 

\newcommand\arcdeg{\mbox{$^\circ$}}%
\newcommand\arcmin{\mbox{$^\prime$}}%
\newcommand\fs{\mbox{$.\!\!^{\mathrm s}$}}%
\newcommand\farcs{\mbox{$.\!\!^{\prime\prime}$}}%

\usepackage[user,hyperref]{zref}

\makeatletter
\providecommand{\@currentshorttitle}{}
\zref@newprop{shorttitle}{\@currentshorttitle}
\zref@addprop{main}{shorttitle}

\NewDocumentCommand{\labelshort}{om}{%
  \begingroup
  \IfValueT{#1}{%
    \renewcommand{\@currentshorttitle}{#1}%
    \zlabel{#2}%
  }%
  \endgroup
  \label{#2}%
}

\NewDocumentCommand{\nameshortref}{O{}m}{%
  \zref@ifrefundefined{#2}{%
  }{%
    \hyperlink{\zref@extract{#2}{anchor}}{#1\zref@extract{#2}{shorttitle}}%
  }%
}

\newcolumntype{C}[1]{>{\centering}m{#1}}

\begin{document}

\title{A cosmic formation site of silicon and sulphur revealed by a new type of supernova explosion}

\author*[1, 2]{\fnm{Steve} \sur{Schulze}}\email{steve.schulze@northwestern.edu}
\author[3]{\fnm{Avishay} \sur{Gal-Yam}}
\author[4]{\fnm{Luc} \sur{Dessart}}
\author[5, 1]{\fnm{Adam A.} \sur{Miller}}
\author[6]{\fnm{Stan E.} \sur{Woosley}}
\author[7]{\fnm{Yi} \sur{Yang} \begin{CJK}{UTF8}{gbsn}
(杨轶)
\end{CJK}}
\author[8, 9, 10]{\fnm{Mattia} \sur{Bulla}}
\author[3]{\fnm{Ofer} \sur{Yaron}}
\author[11]{\fnm{Jesper} \sur{Sollerman}}
\author[12]{\fnm{Alexei V.} \sur{Filippenko}}
\author[13]{\fnm{K-Ryan} \sur{Hinds}}
\author[13]{\fnm{Daniel A.} \sur{Perley}}
\author[14, 15]{\fnm{Daichi} \sur{Tsuna}}
\author[11]{\fnm{Ragnhild} \sur{Lunnan}}
\author[2, 16]{\fnm{Nikhil} \sur{Sarin}}
\author[11]{\fnm{Se{\'a}n J.} \sur{Brennan}}
\author[12]{\fnm{Thomas G.} \sur{Brink}}
\author[17]{\fnm{Rachel J.} \sur{Bruch}}
\author[3]{\fnm{Ping} \sur{Chen}}
\author[18]{\fnm{Kaustav K.} \sur{Das}}
\author[19]{\fnm{Suhail} \sur{Dhawan}}
\author[11]{\fnm{Claes} \sur{Fransson}}
\author[20, 14]{\fnm{Christoffer} \sur{Fremling}}
\author[11]{\fnm{Anjasha} \sur{Gangopadhyay}}
\author[3]{\fnm{Ido} \sur{Irani}}
\author[11]{\fnm{Anders} \sur{Jerkstrand}}
\author[21,22]{\fnm{Nikola} \sur{Kne\v zevi\'c}}
\author[3]{\fnm{Doron} \sur{Kushnir}}
\author[23]{\fnm{Keiichi} \sur{Maeda}}
\author[24]{\fnm{Kate} \sur{Maguire}}
\author[3]{\fnm{Eran} \sur{Ofek}}
\author[13]{\fnm{Conor M. B.} \sur{Omand}}
\author[14]{\fnm{Yu-Jing} \sur{Qin}}
\author[14]{\fnm{Yashvi} \sur{Sharma}}
\author[25]{\fnm{Tawny} \sur{Sit}}
\author[26,27,28]{\fnm{Gokul P.} \sur{Srinivasaragavan}}
\author[3]{\fnm{Nora L.} \sur{Strothjohann}}
\author[29,15,30]{\fnm{Yuki} \sur{Takei}}
\author[3]{\fnm{Eli} \sur{Waxman}}
\author[19]{\fnm{Lin} \sur{Yan}}
\author[31, 12]{\fnm{Yuhan} \sur{Yao}}
\author[12]{\fnm{WeiKang} \sur{Zheng}}
\author[3]{\fnm{Erez A.} \sur{Zimmerman}}
\author[32]{\fnm{Eric C.} \sur{Bellm}}
\author[33]{\fnm{Michael~W.} \sur{Coughlin}}
\author[34]{\fnm{Frank.~J.} \sur{Masci}}
\author[20]{\fnm{Josiah} \sur{Purdum}}
\author[35]{\fnm{Micka\"el} \sur{Rigault}}
\author[34]{\fnm{Avery} \sur{Wold}}
\author[14]{\fnm{Shrinivas~R.} \sur{Kulkarni}}

\affil[1]{\orgdiv{Center for Interdisciplinary Exploration and Research in Astrophysics (CIERA)}, \orgname{Northwestern University}, \orgaddress{\street{1800 Sherman Ave}, \city{Evanston}, \state{IL} \postcode{60201}, \country{USA}}}
\affil[2]{\orgdiv{The Oskar Klein Centre, Department of Physics}, \orgname{Stockholm University}, \orgaddress{\street{AlbaNova University Center}, \city{Stockholm}, \postcode{106 91}, \country{Sweden}}}
\affil[3]{\orgdiv{Department of Particle Physics and Astrophysics}, \orgname{Weizmann Institute of Science}, \orgaddress{\street{234 Herzl St}, \city{Rehovot}, \postcode{76100}, \country{Israel}}}
\affil[4]{\orgdiv{Institut d'Astrophysique de Paris}, \orgname{CNRS-Sorbonne Universit\'e}, \orgaddress{\street{98 bis boulevard Arago}, \city{Paris}, \postcode{75014}, \country{France}}}
\affil[5]{\orgdiv{Department of Physics and Astronomy}, \orgname{Northwestern University}, \orgaddress{\street{2145 Sheridan Rd}, \city{Evanston}, \state{IL} \postcode{60208}, \country{USA}}}
\affil[6]{\orgdiv{Department of Astronomy and Astrophysics}, \orgname{University of California}, \orgaddress{\city{Santa Cruz}, \state{CA} \postcode{95064}, \country{USA}}}
\affil[7]{\orgdiv{Physics Department}, \orgname{Tsinghua University}, \orgaddress{\city{Beijing}, \postcode{100084}, \country{People's Republic of China}}}
\affil[8]{\orgdiv{Department of Physics and Earth Science}, \orgname{University of Ferrara}, \orgaddress{\street{via Saragat 1}, \city{Ferrara}, \postcode{44122}, \country{Italy}}}
\affil[9]{\orgname{INFN, Sezione di Ferrara}, \orgaddress{\street{via Saragat 1}, \city{Ferrara}, \postcode{44122}, \country{Italy}}}
\affil[10]{\orgname{INAF, Osservatorio Astronomico d'Abruzzo}, \orgaddress{\street{via Mentore Maggini snc}, \city{Teramo}, \postcode{64100}, \country{Italy}}}
\affil[11]{\orgdiv{The Oskar Klein Centre, Department of Astronomy}, \orgname{Stockholm University}, \orgaddress{\street{AlbaNova University Center}, \city{Stockholm}, \postcode{106 91}, \country{Sweden}}}
\affil[12]{\orgdiv{Department of Astronomy}, \orgname{University of California}, \orgaddress{\street{501 Campbell Hall}, \city{Berkeley}, \state{CA} \postcode{94720-3411}, \country{USA}}}
\affil[13]{\orgdiv{Astrophysics Research Institute}, \orgname{Liverpool John Moores University}, \orgaddress{\street{Liverpool Science Park IC2, 146 Brownlow Hill}, \city{Liverpool}, \postcode{L3 5RF}, \country{UK}}}
\affil[14]{\orgdiv{Division of Physics, Mathematics and Astronomy}, \orgname{California Institute of Technology}, \orgaddress{\city{Pasadena}, \state{CA} \postcode{91125}, \country{USA}}}
\affil[15]{\orgdiv{Research Center for the Early Universe, School of Science}, \orgname{The University of Tokyo}, \orgaddress{\street{7-3-1 Hongo, Bunkyo-ku}, \city{Tokyo} \postcode{113-0033}, \state{Japan}}}
\affil[16]{\orgdiv{Nordita}, \orgname{Stockholm University and KTH Royal Institute of Technology}, \orgaddress{\street{Hannes Alfv\'ens v\"ag 12}, \postcode{106 91}, \city{Stockholm}, \country{Sweden}}}
\affil[17]{\orgdiv{School of Physics and Astronomy}, \orgaddress{\orgname{Tel Aviv University}, \city{Tel Aviv}, \postcode{69978}, \country{Israel}}}
\affil[18]{\orgdiv{Cahill Center for Astrophysics}, \orgname{California Institute of Technology}, \orgaddress{\street{MC 249-17, 1200 East California Blvd}, \city{Pasadena}, \state{CA} \postcode{91125}, \country{USA}}}
\affil[19]{\orgdiv{Institute of Astronomy and Kavli Institute for Cosmology}, \orgname{University of Cambridge}, \orgaddress{\street{Madingley Road}, \city{Cambridge}, \postcode{CB3 0HA}, \country{UK}}}
\affil[20]{\orgdiv{Caltech Optical Observatories}, \orgname{California Institute of Technology}, \orgaddress{\street{1200 E California Blvd}, \city{Pasadena}, \state{CA} \postcode{91125}, \country{USA}}}
\affil[21]{\orgname{Astronomical Observatory}, \orgaddress{\street{Volgina 7}, \postcode{11060} \city{Belgrade}, \country{Serbia}}}
\affil[22]{\orgname{Department of Astronomy, Faculty of Mathematics}, \orgname{University of Belgrade}, \orgaddress{\street{Studentski trg 16}, \postcode{11000} \city{Belgrade}, \country{Serbia}}}
\affil[23]{\orgdiv{Department of Astronomy}, \orgname{Kyoto University}, \orgaddress{\street{Kitashirakawa-Oiwake-cho, Sakyo-ku}, \city{Kyoto}, \state{Kyoto} \postcode{606-8502}, \country{Japan}}}
\affil[24]{\orgdiv{School of Physics}, \orgname{Trinity College Dublin}, \orgaddress{\street{The University of Dublin}, \city{Dublin}, \postcode{2}, \country{Ireland}}}
\affil[25]{\orgdiv{Department of Astronomy}, \orgname{The Ohio State University}, \orgaddress{\city{Columbus}, \state{OH} \postcode{43210}, \country{USA}}}
\affil[26]{\orgdiv{Department of Astronomy}, \orgname{University of Maryland}, \orgaddress{\city{College Park}, \state{MD} \postcode{20742}, \country{USA}}}
\affil[27]{\orgdiv{Joint Space-Science Institute}, \orgname{University of Maryland}, \orgaddress{\city{College Park}, \state{MD} \postcode{20742}, \country{USA}}}
\affil[28]{\orgdiv{Astrophysics Science Division}, \orgname{NASA Goddard Space Flight Center}, \orgaddress{\street{8800 Greenbelt Rd}, \city{Greenbelt}, \state{MD} \postcode{20771}, \country{USA}}}
\affil[29]{\orgdiv{Yukawa Institute for Theoretical Physics}, \orgname{Kyoto University}, \orgaddress{\street{Kitashirakawa-Oiwake-cho, Sakyo-ku}, \city{Kyoto}, \state{Kyoto} \postcode{606-8502}, \country{Japan}}}
\affil[30]{\orgdiv{Astrophysical Big Bang Laboratory}, \orgname{RIKEN}, \orgaddress{\street{2-1 Hirosawa}, \city{Wako}, \state{Saitama} \postcode{351-0198}, \country{Japan}}}
\affil[31]{\orgname{Miller Institute for Basic Research in Science}, \orgaddress{\street{468 Donner Lab}, \city{Berkeley}, \state{CA} \postcode{94720}, \country{USA}}}
\affil[32]{\orgdiv{DIRAC Institute, Department of Astronomy}, \orgname{University of Washington}, \orgaddress{\street{3910 15th Avenue NE}, \city{Seattle}, \state{WA} \postcode{98195}, \country{USA}}}
\affil[33]{\orgdiv{School of Physics and Astronomy}, \orgname{University of Minnesota}, \orgaddress{\street{116 Church Street S.E.}, \city{Minneapolis}, \state{MN} \postcode{55455}, \country{USA}}}
\affil[34]{\orgdiv{IPAC}, \orgname{California Institute of Technology}, \orgaddress{\street{1200 E. California Blvd}, \city{Pasadena}, \state{CA} \postcode{91125}, \country{USA}}}
\affil[35]{\orgdiv{IP2I Lyon / IN2P3, IMR 5822}, \orgname{Universite Claude Bernard Lyon 1, CNRS}, \orgaddress{\street{Enrico Fermi}, \city{Villeurbanne}, \postcode{69622}, \country{France}}}

\abstract{
The cores of stars are the cosmic furnaces where light elements are fused into heavier nuclei \cite{Burbidge1957a, Kippenhahn2013a, Arcones2023a}. The fusion of hydrogen to helium initially powers all stars. The ashes of the fusion reactions are then predicted to serve as fuel in a series of stages, eventually transforming massive stars into a structure of concentric shells. These are composed of natal hydrogen on the outside, and consecutively heavier compositions inside, predicted to be dominated by helium, carbon/oxygen, oxygen/neon/magnesium, and oxygen/silicon/sulphur \cite{Woosley1995a, Woosley2002a}. Silicon and sulphur are fused into inert iron, leading to the collapse of the core and either a supernova explosion or the direct formation of a black hole \cite{Heger2003a, Woosley2005a,Mueller2016a,Woosley2017a}. Stripped stars, where the outer hydrogen layer has been removed and the internal He-rich layer (in Wolf-Rayet WN stars) or even the C/O layer below it (in Wolf-Rayet WC/WO stars) are exposed \cite{Crowther2007a}, provide evidence for this shell structure, and the cosmic element production mechanism it reflects. The types of supernova explosions that arise from stripped stars embedded in shells of circumstellar material (most notably Type Ibn supernovae from stars with outer He layers, and Type Icn supernovae from stars with outer C/O layers) confirm this scenario \cite{Matheson2000b, Pastorello2007a, Gal-Yam2022a, Perley2022a, Maeda2022a}. However, direct evidence for the most interior shells, which are responsible for the production of elements heavier than oxygen, is lacking. Here, we report the discovery of the first-of-its-kind supernova arising from a star peculiarly stripped all the way to the silicon and sulphur-rich internal layer. Whereas the concentric shell structure of massive stars is not under debate, it is the first time that such a thick, massive silicon and sulphur-rich shell, expelled by the progenitor shortly before the SN explosion, has been directly revealed.
}

\maketitle

\begin{figure}
    \centering
    \includegraphics[width=1\textwidth]{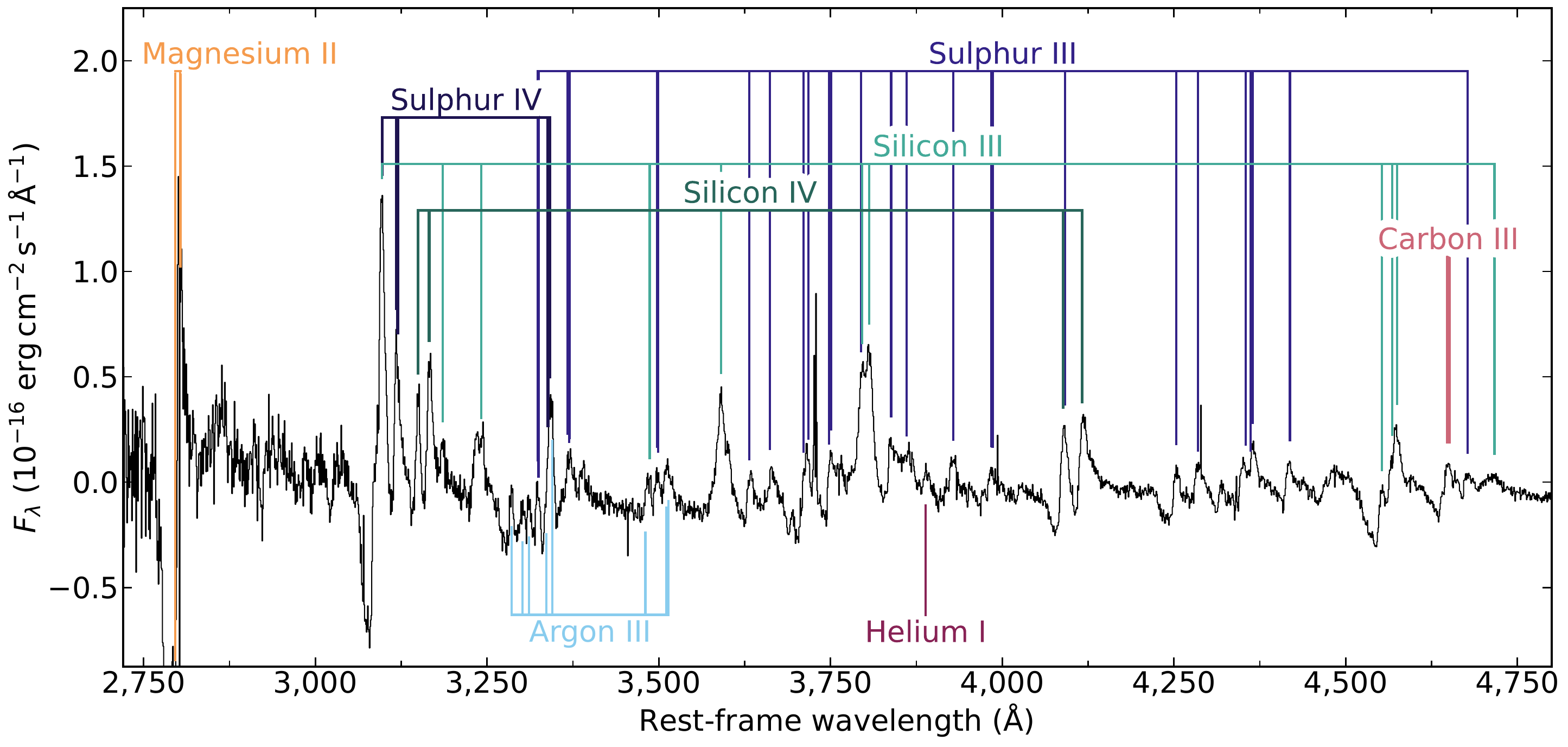}
    \caption{
    \textbf{Spectrum obtained 1.0 days after the first ZTF detection with Keck/LRIS in the range from 2,720 to 4,800 \AA, after subtracting the blackbody continuum.} The spectrum reveals narrow emission lines of highly ionised species of silicon, sulphur, and argon, which have never been seen in any SN before, as well as doubly ionised carbon, singly ionised magnesium, and neutral helium. A number of the highly ionised silicon and sulphur lines also exhibit P~Cygni profiles with a maximum velocity of $\sim 3{,}000~\rm km\,s^{-1}$ (\extmat \figuretwo \ref{fig:spec:line_profiles}), indicating that these lines are produced in a fast-moving, metal-rich CSM (e.g., a wind or a shell expelled by the progenitor shortly before the explosion). The spectrum is rebinned for illustration purposes. The lower bound of the ordinate axis was cropped for illustration purposes. 
    }
    \label{fig:spec:flash:id}
\end{figure}

On 7 September 2021 at 09:56 (UTC dates are used throughout this paper), the public Northern Sky Survey of the Zwicky Transient Facility (ZTF) \cite{Bellm2019a, Graham2019a} discovered the supernova (SN) 2021yfj at right ascension $\alpha=01^{\rm h}37^{\rm m}46\fs171$ and declination $\delta=-01\arcdeg15'17\farcs78$ (J2000.0; \method \nameref{app:discovery}) \cite[][]{Munoz-Arancibia2021a}. A spectrum obtained with Keck/LRIS 24 hours after discovery shows a large number of narrow emission lines and P~Cygni profiles from ionised silicon, sulphur, and argon (\ion{Si}{iii}--\textsc{iv}, \ion{S}{iii}--\textsc{iv}, and \ion{Ar}{iii}; \figuretwo \ref{fig:spec:flash:id}) superimposed on a hot blackbody spectrum ($T\approx15{,}000~\rm K$; \method \nameref{app:lc:bolometrics}), previously unobserved in any supernova \cite{Bruch2023a, Gal-Yam2022a, Pastorello2008a}. Lines of lighter elements, which are much more common in the Universe \cite{Arcones2023a} and usually detected in spectra of infant SNe \cite{Bruch2021a, Gal-Yam2022a, JacobsonGalan2024a}, are either very weak (carbon and helium) or even completely absent (e.g., hydrogen, nitrogen). This prompted us to monitor the photometric and spectroscopic evolution at optical and ultraviolet (UV) wavelengths for the next 120 days until \sn faded below the brightness of the host galaxy (\suppmat \nameref{app:obs}). The early spectra also reveal narrow absorption and emission lines from the interstellar medium (ISM) in the host galaxy at a redshift of $z=0.13865\pm0.00004$, placing \sn at a luminosity distance of $\approx676.4$~Mpc, assuming Planck cosmology (\method \nameref{app:redshift}; \cite{Planck2018a}). Hereafter, all times are given with respect to the discovery time and in the rest-frame.

The \ion{Si}{iii}--\textsc{iv}, \ion{S}{iii}--\textsc{iv}, and \ion{Ar}{iii} lines are visible up to day 7.7 (i.e., up to $\approx5.4$ days after the time of the $g$-band maximum). With time, these lines become weaker, and emission lines from singly-ionised silicon and sulphur emerge (\extmat \figuretwo \ref{fig:spec:sequence}). Some lines exhibit P~Cygni profiles (\extmat \figuretwo \ref{fig:spec:line_profiles}), which can be produced either in the SN ejecta, an expanding shell of gas expelled by the progenitor prior to the explosion, or a stellar wind. The absorption minima are at a velocity of 1,000--1,500~$\rm km\,s^{-1}$ and the blue edge of the absorption component reaches merely $3{,}000~\rm km\,s^{-1}$. This is significantly slower than typical SN ejecta velocities ($\sim10{,}000~\rm km\,s^{-1}$; \cite[][]{Liu2017a}), but comparable to the wind velocities of Wolf-Rayet (WR) stars \cite{Crowther2007a} and similar to the velocities of expanding shells of circumstellar material (CSM) around some SNe \cite[e.g.,][]{Matheson2000b, Pastorello2007a, Gal-Yam2022a, Perley2022a}. Between day 11 and 20, the blackbody spectrum subsides, and a blue pseudocontinuum dominates the emission in the optical, a tell-tale sign of interaction with CSM \cite{Dessart2022a}. Superimposed are emission lines from \ion{Si}{ii}, \ion{S}{ii}, \ion{Mg}{i}--\textsc{ii}, \ion{He}{i}, and \ion{O}{i} that remain visible through our last spectrum at day 49.8 (\extmat \figuretwo \ref{fig:spec:sequence}). Considering that silicon, sulphur, and argon are the ashes of the ephemeral oxygen-burning phase that takes place less than a few years before a massive star dies \cite[e.g.,][]{Woosley2002a}, it is puzzling that the early and late spectra also show helium (\method \nameref{app:spec:helium}). Helium should have been consumed during the earlier burning stages, and it is not a daughter product of the oxygen-burning phase \cite{Woosley2002a}.

\begin{figure}
    \centering
    \includegraphics[width=0.75\textwidth]{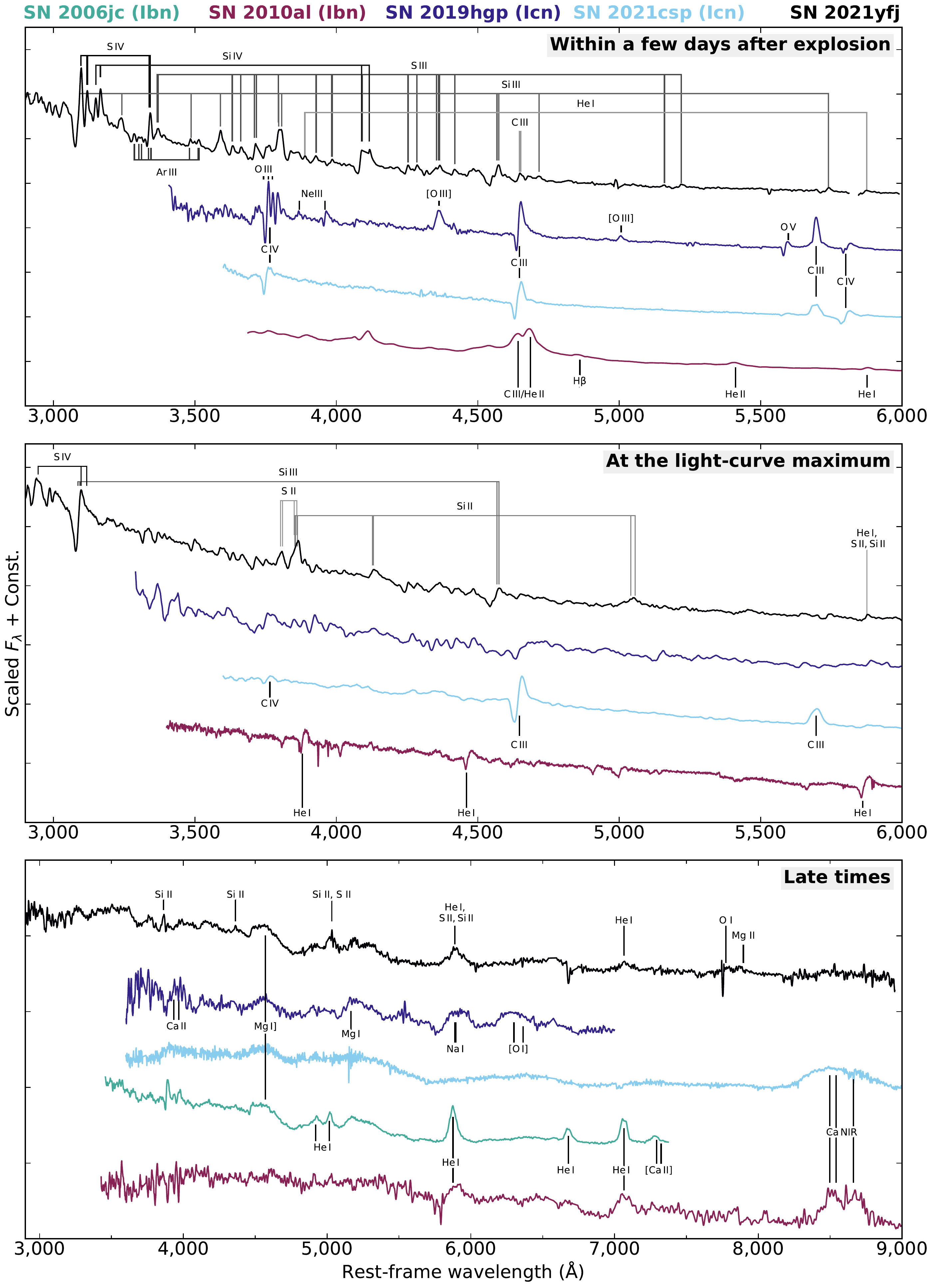}
    \caption{\textbf{Comparison of \sn with other interaction-powered SESNe a few days after the explosion (top), the time of maximum light (middle), and late times (bottom).} The earliest spectra of Type Ibn and Icn SNe are characterised by a hot blackbody continuum with a temperature similar to that of \sn 
    (\extmat \figuretwo \ref{fig:lc:comparison}). They also show lines of helium, carbon and possibly hydrogen (in SN 2010al) (Ibn), or carbon, oxygen, and neon (Icn), but no silicon, sulphur, and argon. These strong helium, carbon, oxygen, and neon lines are clearly absent in \sn. Well after peak brightness, the spectra of SNe~Ibn,  Icn, and \sn are characterised by a blue pseudocontinuum with superimposed intermediate-width (a few 1,000~$\rm km~s^{-1}$) emission lines, due to the interaction of the SN ejecta and CSM. Again, \sn shows prominent silicon and sulphur emission lines that are absent in the other objects, and the comparison objects exhibit features that are clearly absent in \sn. Therefore, the differences between \sn and previously known classes of interaction-powered SESNe reflect true differences in the CSM composition and the progenitor populations, making \sn the first member of a previously unknown supernova class. 
    All spectra are rebinned for illustration purposes.
    }
    \label{fig:spec:ibn_comparison}
\end{figure}

\sn's unique properties vividly stand out when comparing its spectroscopic sequence to those of stripped-envelope supernovae (SESNe) that strongly interact with CSM rich in either He (Type Ibn) or C/O/Ne (Type Icn; \figuretwo \ref{fig:spec:ibn_comparison}, \method \nameref{app:spec:comparison}). Spectra of SNe~Ibn and Icn obtained shortly after explosion exhibit narrow emission lines from carbon, nitrogen, oxygen, helium, or hydrogen, but no silicon and no sulphur. These differences remain well visible in the spectra obtained around maximum light. The Type Icn SNe 2019hgp and 2021csp also show P~Cygni profiles with velocities similar to those of \sn, which were interpreted as originating from a fast stellar wind \cite{Gal-Yam2022a, Perley2022a}. At late times, all objects are characterised by a blue pseudocontinuum with superimposed emission lines. Yet again, \sn displays striking differences: conspicuous silicon and sulphur emission lines but no Ca near-infrared (NIR) triplet at $\sim8{,}500$~\AA\ in emission (\figuretwo \ref{fig:spec:ibn_comparison}), whereas the other objects show \ion{He}{i} (SNe Ibn) and a conspicuous Ca NIR triplet in emission. Therefore, the discovery of silicon and sulphur in \sn is not due to differences in the data quality, the timing of the observations, or the blackbody temperature. Instead, the presence of silicon and sulphur reflects true differences in the elemental composition of the CSM and, therefore, the SN progenitor. This lets us conclude that \sn is embedded in a thick, extended CSM rich in silicon and sulphur, never observed in any supernova before. Since SN classes are defined by the absence or presence of particular chemical elements in their spectra \citep{Filippenko1997a, Gal-Yam2017a}, \sn is the first member of a previously unknown SN class: Type Ien (\method \nameref{app:newsntype}; \cite{Gal-Yam2024a}).

\begin{figure}
    \centering
    \includegraphics[width=0.5\textwidth]{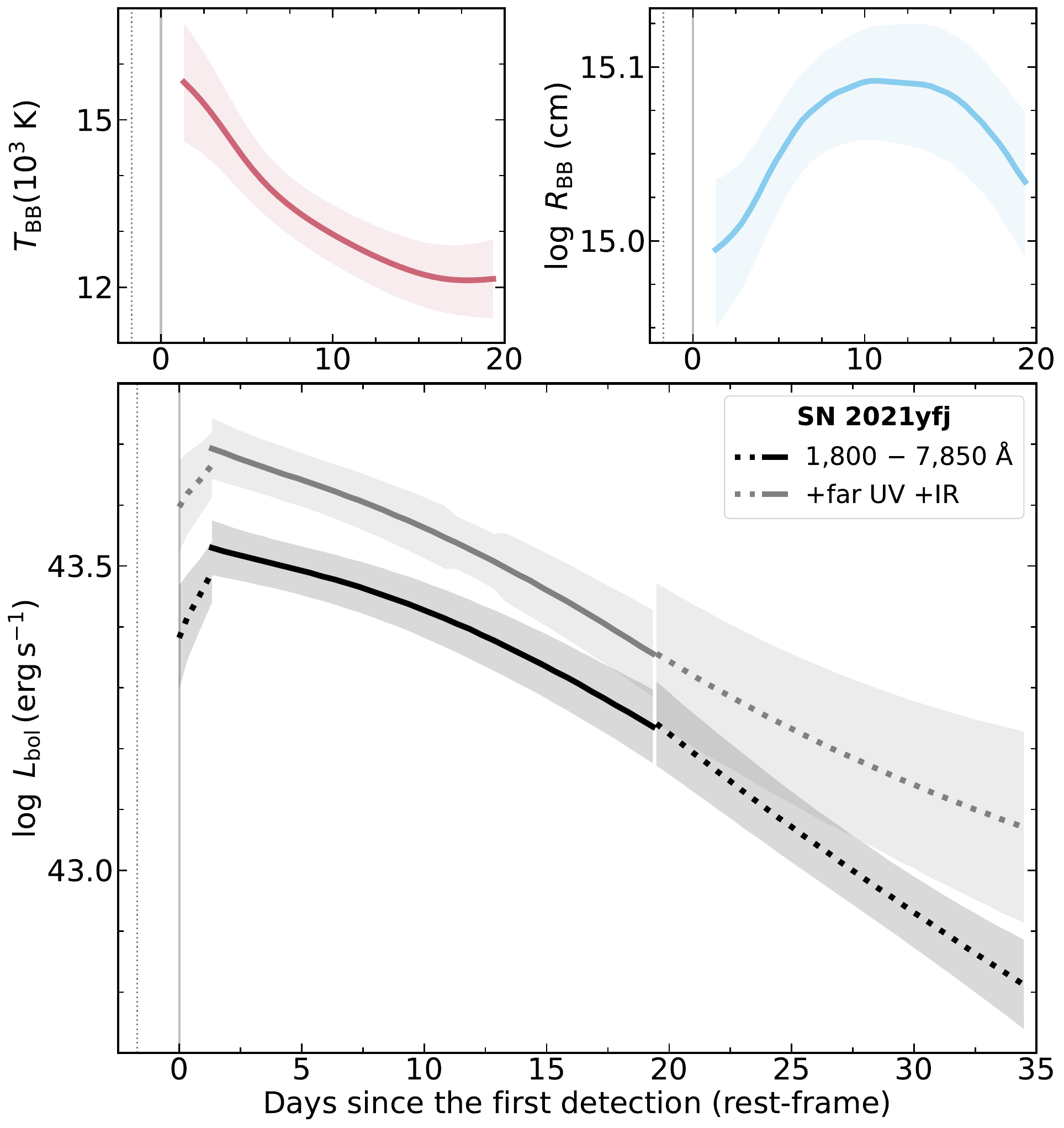}
    \caption{\textbf{The bolometric light curve of \sn (bottom panel) and the evolution of the blackbody temperature (top right) and radius (top left).}
    The bolometric light curve shown in black covers the wavelength interval from 1,800 to 7,850 \AA. Correcting the bolometric flux for the missing far-UV and IR flux increases the luminosity by $\sim0.2$~dex. At peak brightness, \sn reached a luminosity of $>3\times10^{43}~\rm erg\,s^{-1}$. Integrating over the entire light curve yields a radiated energy of (0.6--1) $\times10^{50}~\rm erg$. The smaller value covers the range from 1,800 to 7,850~\AA, and the larger value includes an estimate of the contribution from the far-UV and IR.
    Dotted lines indicate time intervals with incomplete wavelength coverage. The blackbody temperature and radius have typical values for infant interaction-powered SNe, although the gradual evolution of both properties is atypical for infant supernovae and indicative of an optically thick CSM (\method \nameref{app:lc:bolometrics}).
    The shaded bands indicate the statistical uncertainties at the $1\sigma$ confidence level. The vertical dotted line in each panel indicates the date of the last non-detection.}
    \label{fig:lc:bolometric}
\end{figure}

\sn is $>2$ mag more luminous than a typical SESN ($M_g=-19.0$~mag for \sn compared to $M_g=-17.3$~mag for a typical SESN), and it has a shorter rise time from 50\% peak flux (2.3~days vs. 8 days). The combination of being more luminous with a shorter rise time is inconsistent with Ni-powered SNe. The modelling of the light curve corroborates this. Solely powering by radioactive $^{56}$Ni or a central engine, such as a spindown of a rapidly spinning, highly magnetised neutron star, can be excluded (\method \nameref{app:lc:modelling}). Owing to the assumptions of the two models, they do not simultaneously capture the rise, peak luminosity, and decline. The bolometric light curve (\figuretwo \ref{fig:lc:bolometric}), constructed from the multiband data is akin to those of interaction-powered SNe~Ibn and Icn. At peak, \sn reached $\gtrsim3\times10^{43}~\rm erg\,s^{-1}$, and integrating over the entire light curve yields a radiated energy of 0.6--$\sim1\times10^{50}~\rm erg$. Both values are strict lower limits because the true bolometric peak likely occurred before our multiwavelength campaign started (\method \nameref{app:lc:bolometrics}). Fitting the bolometric light curve with the model of ejecta-CSM interaction from Ref. \cite{Takei2022a, Takei2024a} reproduces the photometric evolution of \sn (\method \nameref{app:lc:modelling}). The fit points to an ejecta mass of $\sim 5~M_\odot$, a CSM mass of $>1~M_\odot$ and $<M_{\rm ejecta}$, and an explosion energy of (1.5--2) $ \times10^{51}~\rm erg$ (\extmat \figuretwo \ref{fig:lc:redback}). The slight tension between the model and the observations during the rise could be mitigated by more complex CSM geometries and density profiles than the ones used here. 

\begin{figure}
    \centering
    \includegraphics[width=0.725\textwidth]{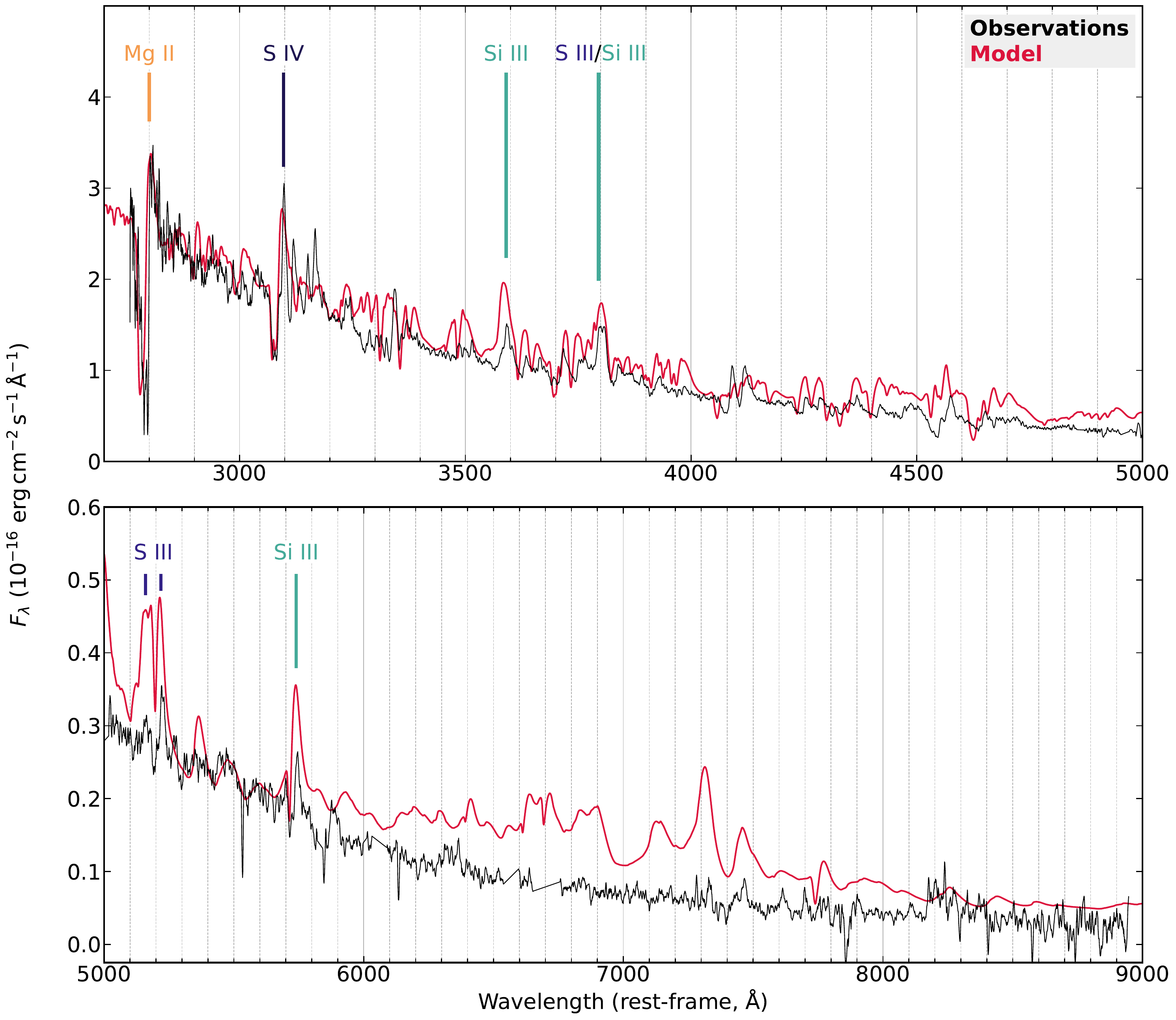}
    \caption{\textbf{Comparison of the discovery spectrum of \sn (black; \figuretwo \ref{fig:spec:flash:id}) with a spectral model (red).} The model assumes a power of $3\times10^{43}~\rm erg\,s^{-1}$, an ejecta mass of $3.24~M_\odot$, a velocity of $1000~\rm km\,s^{-1}$, and an elemental composition similar to the O/Si shell in massive helium stars: 0.786 (oxygen), 0.1 (neon), 0.05 (silicon), 0.03 (sulphur), 0.01 (argon), 0.01 (magnesium), 0.001 (calcium), and solar abundance for iron, cobalt and nickel. The spectrum is computed with steady-state, non-local thermodynamic equilibrium radiative transfer models (for more details see the \methods \nameref{app:spectral_modelling}). This model matches the strongest silicon and sulphur lines and also magnesium in terms of absolute and relative strength and line width, corroborating that \sn is likely the explosion of a massive star stripped down to its O/Si shell.
    }
    \label{fig:spec:model}
\end{figure}

The ubiquity of silicon and sulphur lines in the blue part of the optical spectra at early times raises the question of the silicon and sulphur abundance in the spectrum formation region. Would a low abundance be sufficient, as can happen with iron-group elements? (For example, solar-metallicity iron can easily explain the strong blanketing observed in SNe~II at the recombination epoch; see, e.g., observations \cite{Elmhamdi2003a} and models \cite{Dessart2005a,  Dessart2011a, Hillier2019a}.) Or, would a high abundance be necessary, as can happen for helium in SNe~Ib \cite[e.g.,][]{Dessart2020a}? To address these questions, we perform exploratory radiative-transfer simulations using mock ejecta with an elemental composition similar to the O/Si shell in massive helium stars, which is the outermost shell containing a significant amount of freshly nucleosynthesised silicon and sulphur. 
Our model consists of an ejecta with $3.24~M_\odot$, a steep density profile with a power-law exponent of $-10$ (a wind profile is excluded because of the weakness of emission lines), and a base velocity of $1000~\rm  km\,s^{-1}$, yielding a total kinetic energy of a few 10$^{49}$\,erg at that time. Furthermore, we deposit a power of $3\times10^{43}~\rm erg\,s^{-1}$ at a velocity of $1000~\rm km\,s^{-1}$ (\method \nameref{app:spectral_modelling}). This simple model (shown in red in \figuretwo \ref{fig:spec:model}) can explain the observed features (shown in black in the same figure) at $<4{,}800$~\AA\ in the discovery spectrum in terms of the relative and absolute strength as well as the line width. Increasing the mass fraction from a few percent to 30--50\% yields slightly stronger features from silicon and sulphur but in growing tension with the observations (\method \nameref{app:spectral_modelling}). Likewise, decreasing the mass fraction below 1\% gives weaker features and is, therefore, inconsistent with the observations. Based on these simulations, we conclude that a mass fraction of a few $\times 0.01$ suffice to explain the observed features. Furthermore, to have an optically thick CSM out to $10^{15}~\rm cm$ requires a large CSM mass of $\sim3~M_\odot$. This strongly suggests that \sn is the product of a massive star that was stripped all the way to the silicon- and sulphur-rich internal layer prior to explosion. The presence of helium cannot be explained by this model. In fact, no O- or Si-rich material in massive-star models is rich in He. 

In the framework of wind-driven mass loss, the stripping of a star all the way to the O/Si shell is very challenging to explain. Pulsational pair instabilities might accomplish that \cite{Fowler1964a, Barkat1967a, Rakavy1967a, Woosley2017a}. This phenomenon is predicted to occur in stars with an initial mass of 70--140~$M_\odot$, which experience a recurrent instability from producing $e^-e^+$ pairs during the oxygen-burning phase. The interaction between CSM shells can produce luminous transients \cite{Woosley2017a, Leung2019, Marchant2019} with properties qualitatively matching those of \sn. In this scenario, \sn could be the product of collisions between the last shells ejected before the star collapses into a black hole. While this model has appeal, the detection of helium cannot be readily explained and may require, for example, a binary helium star companion. The presence of helium is also an issue for the other massive-star scenarios discussed in the \method \nameref{app:progenitor}, which are motivated by the star-forming host of \sn (\method \nameref{app:host}). A review of literature models involving accretion and burning of helium on the surface of a compact object does not reveal a suitable Si/S-forming scenario (\method \nameref{app:progenitor}). All these models merit additional theoretical studies.

Stellar evolution theory predicts that stars nearing the end of their lives should consist of concentric shells, with hydrogen in the outermost shell and iron at the core \cite{Burbidge1957a, Kippenhahn2013a, Woosley1995a, Woosley2002a, Arcones2023a}. However, direct observations of this shell structure are rare. WR stars experience significant mass loss toward the end of their lives, which can expose their He (WN stars) and C/O (WC/WO stars) shells \cite{Crowther2007a}. Interaction-powered SNe provide an independent view of the shell structure of stars. The CSM they interact with encodes information on the surface composition of the dying star just before it explodes, free from contamination by explosion products \cite{Gal-Yam2014a, Groh2014a, Yaron2017a}. Type IIn, Ibn, and Icn supernovae probe the H, He, and C/O shells, respectively. These SNe not only complement observations of Wolf-Rayet stars but also reveal how dying stars can lose part of their outer layers just before the terminal explosion. The discovery of \sn has three important implications for stellar evolution theory:
(i) it reveals a formation site of argon, silicon, and sulphur;
(ii) it likely directly confirms the complete sequence of concentric shells in massive stars; and
(iii) it requires the operation of processes that can strip stars down to their inner shells.

The lack of known SNe that show any similarity to the early- or late-time spectra of \sn\ suggests that Type Ien SNe are intrinsically rare. The ZTF Bright Transient Survey \cite[BTS;][]{Fremling2020a, Perley2020a}, which aims to spectroscopically classify all ZTF transients that peak brighter than $m = 18.5$~mag and is nearly 100\% complete, has not identified a single \sn-like event during its six years of operation. Thus, from the BTS, we conclude that the rate of \sn-like events is $<30~\rm Gpc^{-3}\,yr^{-1}$ [95\% confidence], $<1/1{,}000$ the rate of SNe\,Ib/c [\method \nameref{app:rates}; \citenum{Li2011a}]. As future facilities, such as the Vera C.\ Rubin Observatory, continue to expand the discovery space for the transient Universe, there is a great deal of hope that these surveys will uncover new classes of explosive events. \sn\ represents one of these rare new transients, but it is important to note that it does not significantly stand out from the population of extragalactic transients based on its photometric evolution alone (\extmat \figuretwo \ref{fig:lc:parameters}). Instead, it is the spectra that uniquely identify \sn\ as belonging to an entirely new SN class: narrow silicon and sulphur lines at early times and a blue pseudocontinuum with silicon and sulphur lines at late times. This highlights the importance of spectroscopic observations and lays plain evidence that even sophisticated artificial-intelligence-powered anomaly-detection algorithms running on light curves from the Rubin Legacy Survey of Space and Time \cite[LSST;][]{LSST2009a} will not be able to recover every new type of transient in the LSST data stream. To detect additional \sn-like events or discover the predicted but yet not discovered Type Id/Idn/Ie SNe (\extmat \nameref{app:newsntype}), efforts are needed that tightly couple high-throughput, medium-resolution spectrographs with long-duration, high-cadence time-domain surveys, which will naturally be provided by existing and future medium-deep surveys \cite{Tonry2011a, Bellm2019a, Smith2020a,  Jones2021a, Steeghs2022a, Ofek2023b, Groot2024a} and the deep Rubin Legacy Survey of Space and Time \cite{LSST2009a}. The discovery of any additional \sn-like objects will have a profound impact on our understanding of their nature.

\clearpage

\clearpage
\input{extended_material}
\clearpage
\input{methods}
\clearpage
\input{supplementary_material}

\end{document}

%% file: extended_material.tex
\setcounter{figure}{0}

\section{Extended material}

\begin{figure}[h!]
    \centering
    \captionsetup{name=\extmattwo Fig.}
    \includegraphics[width=0.75\textwidth]{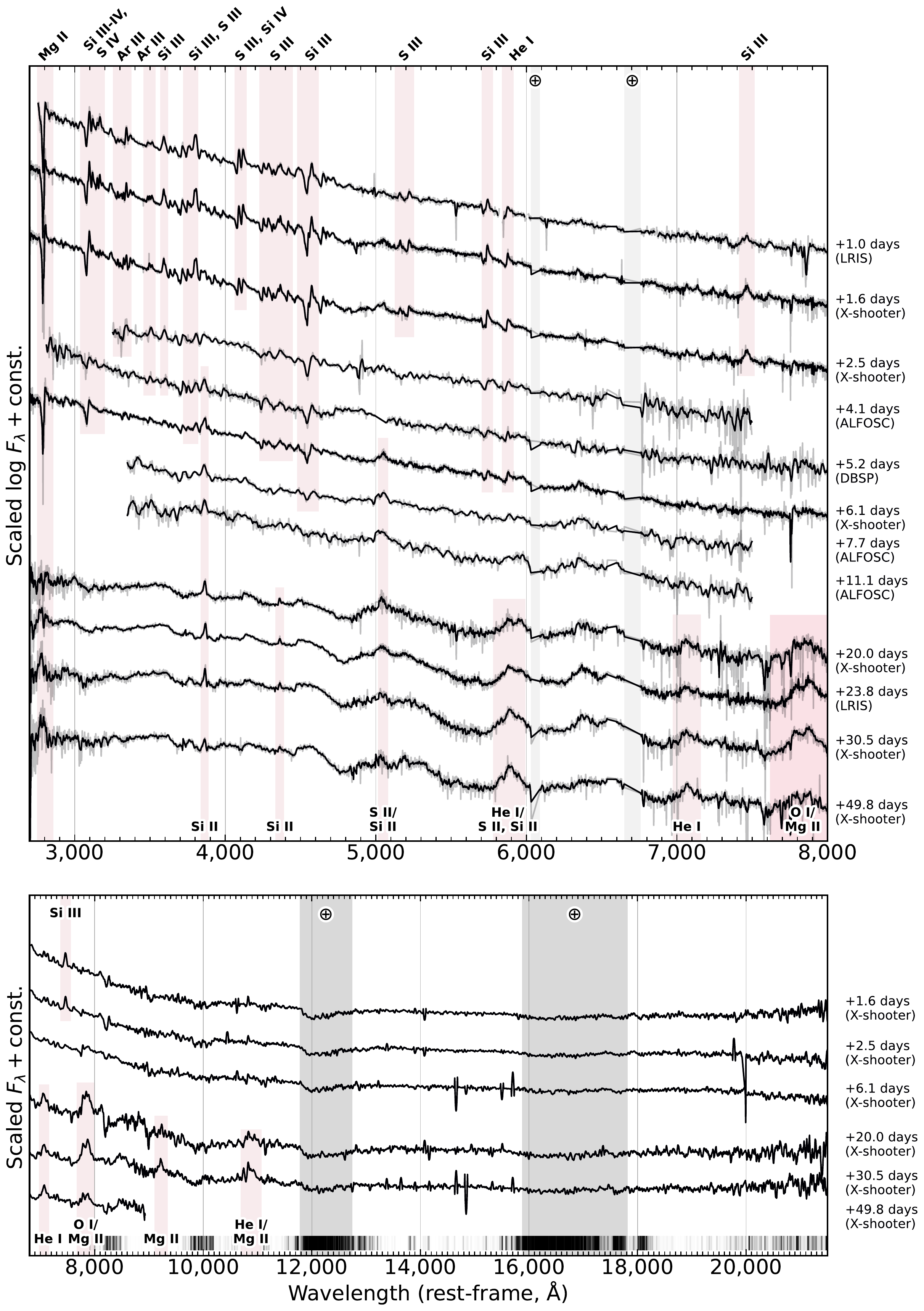}
    \caption{\textbf{Spectral evolution from day 1 to 49.8 of \sn in the UV-optical (upper panel) and near IR (lower panel).} Up to day 11 the spectra are characterised by a blackbody shape with superimposed narrow emission and P~Cygni lines from silicon, sulphur, argon, carbon, and helium. As the photosphere cools, the ionisation state of silicon, sulphur, and argon decreases. By day 20, a blue pseudocontinuum dominates the spectrum with superimposed intermediate-width emission lines from magnesium, silicon, sulphur, and helium. The most prominent features of both phases are marked. Regions of high atmospheric absorption are marked, and a near-IR spectrum of the opacity of Earth's atmosphere is shown as black vertical lines (black = high opacity). Host-galaxy emission lines are clipped. The original spectra are in grey, and rebinned versions are in black.
    }
    \label{fig:spec:sequence}
\end{figure}

\begin{figure}[h!]
    \centering
    \captionsetup{name=\extmattwo Fig.}
    \includegraphics[width=1\textwidth]{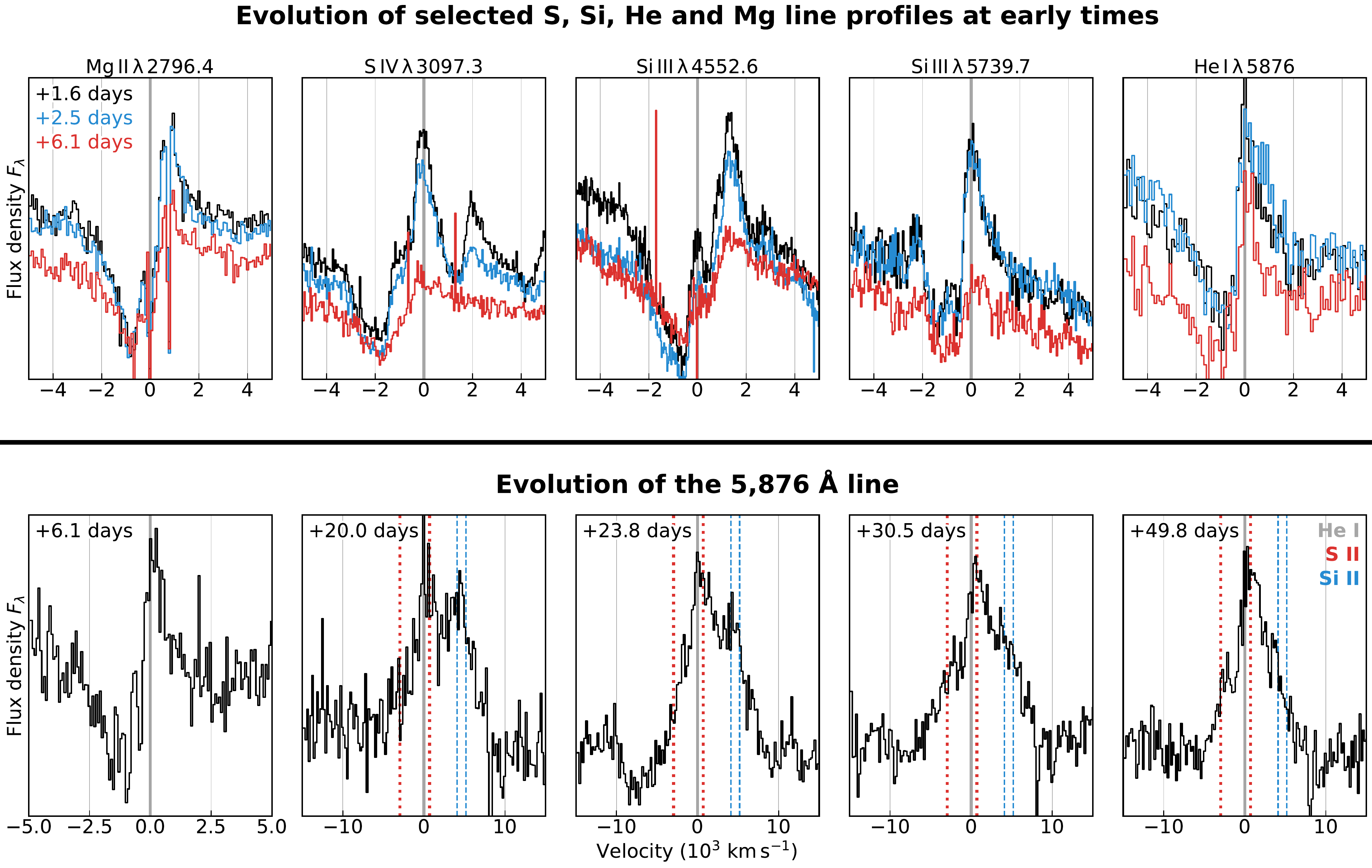}
    \caption{\textbf{The evolution of the line profiles of selected lines from helium, magnesium, silicon, and sulphur.} \textbf{\textit{Top:}} At early times, all lines show well-developed P~Cygni profiles. The absorption minima are at $\sim 1{,}500~\rm km\,s^{-1}$. The blue edge, a proxy of the maximum velocity, extends to $\sim 3{,}000~\rm km\,s^{-1}$. These velocities are comparable to velocities of stellar winds as seen in Wolf-Rayet stars \cite{Crowther2007a} and winds around some SNe Icn \cite{Gal-Yam2022a, Perley2022a}, and much slower than SN-ejecta velocities at similar phases ($\sim10{,}000~\rm km\,s^{-1}$; \cite[][]{Liu2016a}). The \ion{Si}{iii} and \ion{S}{iv} lines are blended with other lines and exhibit complex line profiles. The \ion{Mg}{ii} line shows narrow absorption lines from the ISM in the host galaxy.
    \textbf{\textit{Bottom:}} Up to day 6, the $5{,}876$~\AA\ feature shows a well-developed P~Cygni profile and is dominated by \ion{He}{i}. At later phases, this feature transitions into a pure emission line with time-variable contributions from silicon, sulphur, and helium. The spectra are rebinned for illustration purposes.
    }
    \label{fig:spec:line_profiles}
\end{figure}

\begin{figure}[h!]
    \centering
    \captionsetup{name=\extmattwo Fig.}
    \includegraphics[width=0.725\textwidth]{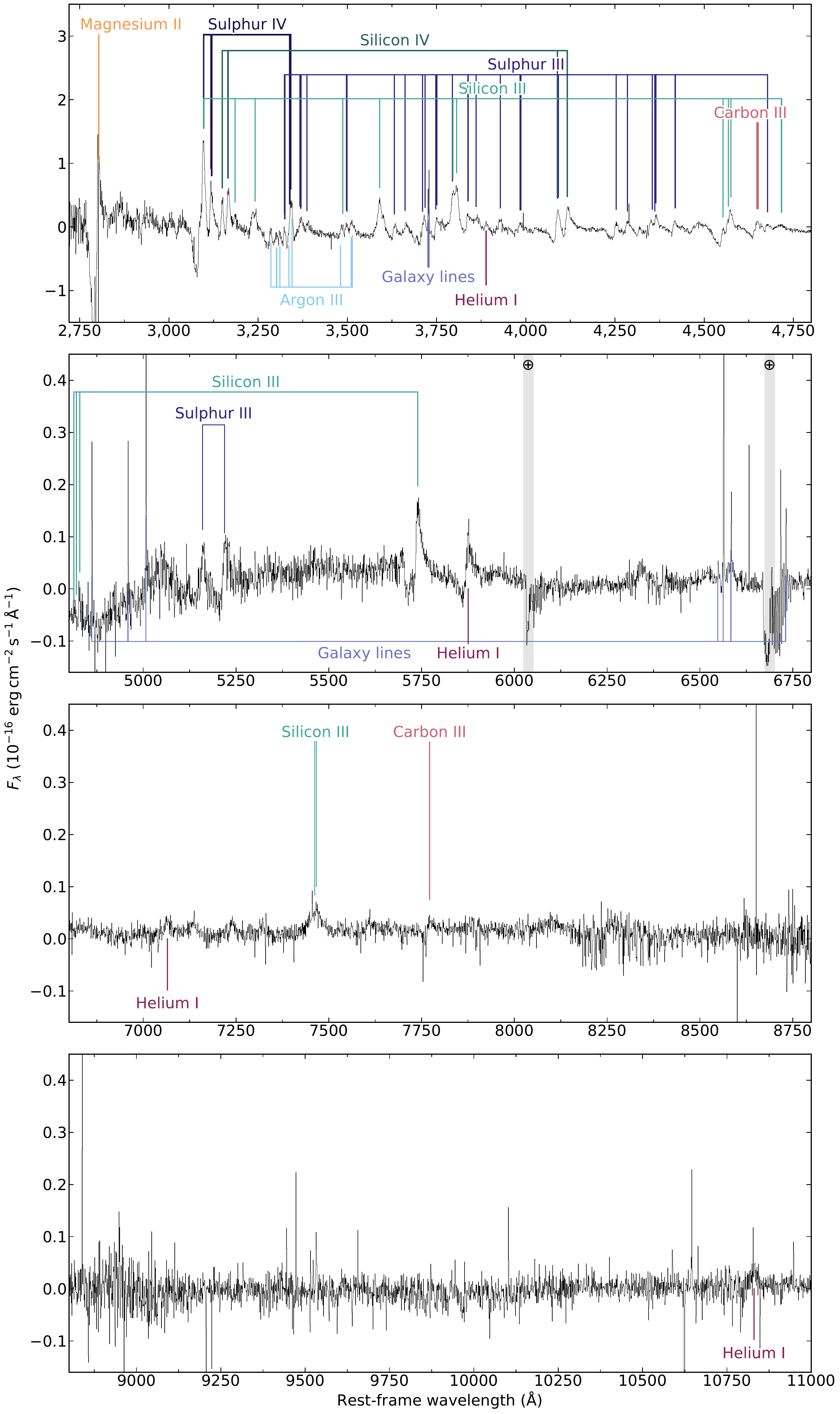}
    \caption{\textbf{Spectrum obtained 1.6 days after the first ZTF detection with VLT/X-shooter, after subtracting the blackbody continuum.} The full spectrum covers the wavelength range from 2,635 to 21,960~\AA. The displayed wavelength range is limited to 2,720--11,000~\AA\ where SN features are well visible. The top panel shows the same wavelength interval as the discovery spectrum in \figuretwo \ref{fig:spec:flash:id} obtained 12 hours earlier. The evolution between both epochs is gradual at most. In addition to the SN features, the spectrum shows emission lines from star-forming regions in the host galaxy and narrow absorption lines from the host ISM. Strong telluric features are marked with ``$\oplus$''.
    }
    \label{fig:spec:flash:id:full}
\end{figure}

\begin{figure}[h!]
    \centering
    \captionsetup{name=\extmattwo Fig.}
    \includegraphics[width=1\textwidth]{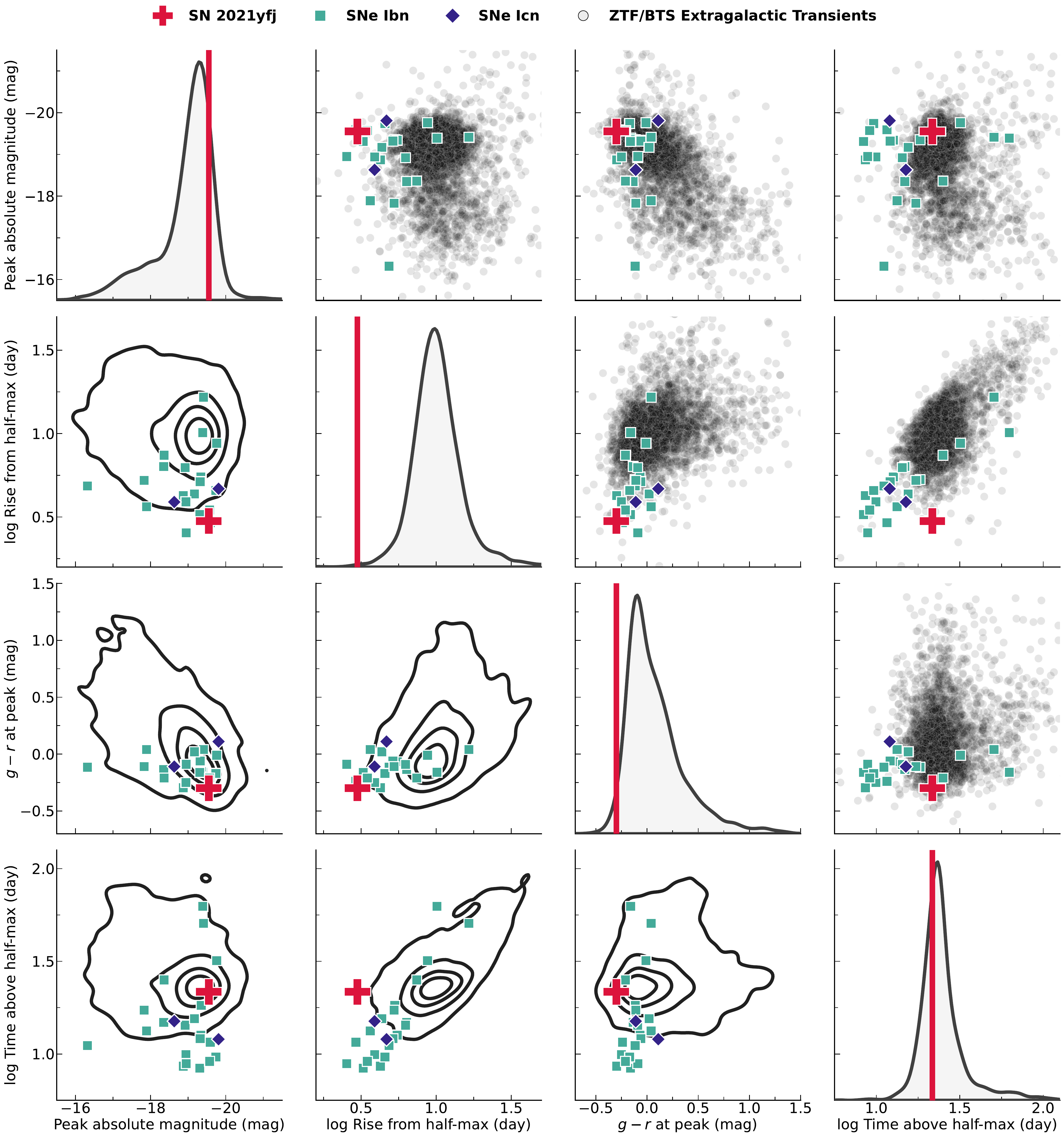}
    \caption{\textbf{\sn in a 4-dimensional light-curve feature space, together with 4032 extragalactic transients from the ZTF Bright Transient Survey (79\% Type Ia SNe, 11\% Type II SNe, and 10\% other types of core-collapse SNe and other types of transients).} The panels above the diagonal show all measurements in different projections of the feature space, the panels below present the diagonal 2-dimensional kernel-density estimates and the panels on the diagonal display 1-dimensional kernel-density estimates. The locations of \sn and Type Ibn/Icn SNe are highlighted in all 2-dimensional projections.    
    \sn's light curve shares similarities with interaction-powered SNe~Ibn and Icn: a fast rise and a high peak luminosity. However, it sustains a high luminosity for a significantly longer period of time, which is uncommon for interaction-powered SESNe but comparable to regular supernovae. The combination of short rise and long duration places \sn in a sparsely populated area of the light-curve parameter space.
    }
    \label{fig:lc:parameters}
\end{figure}

\begin{figure}[h!]
    \centering
    \captionsetup{name=\extmattwo Fig.}
    \includegraphics[width=1\textwidth]{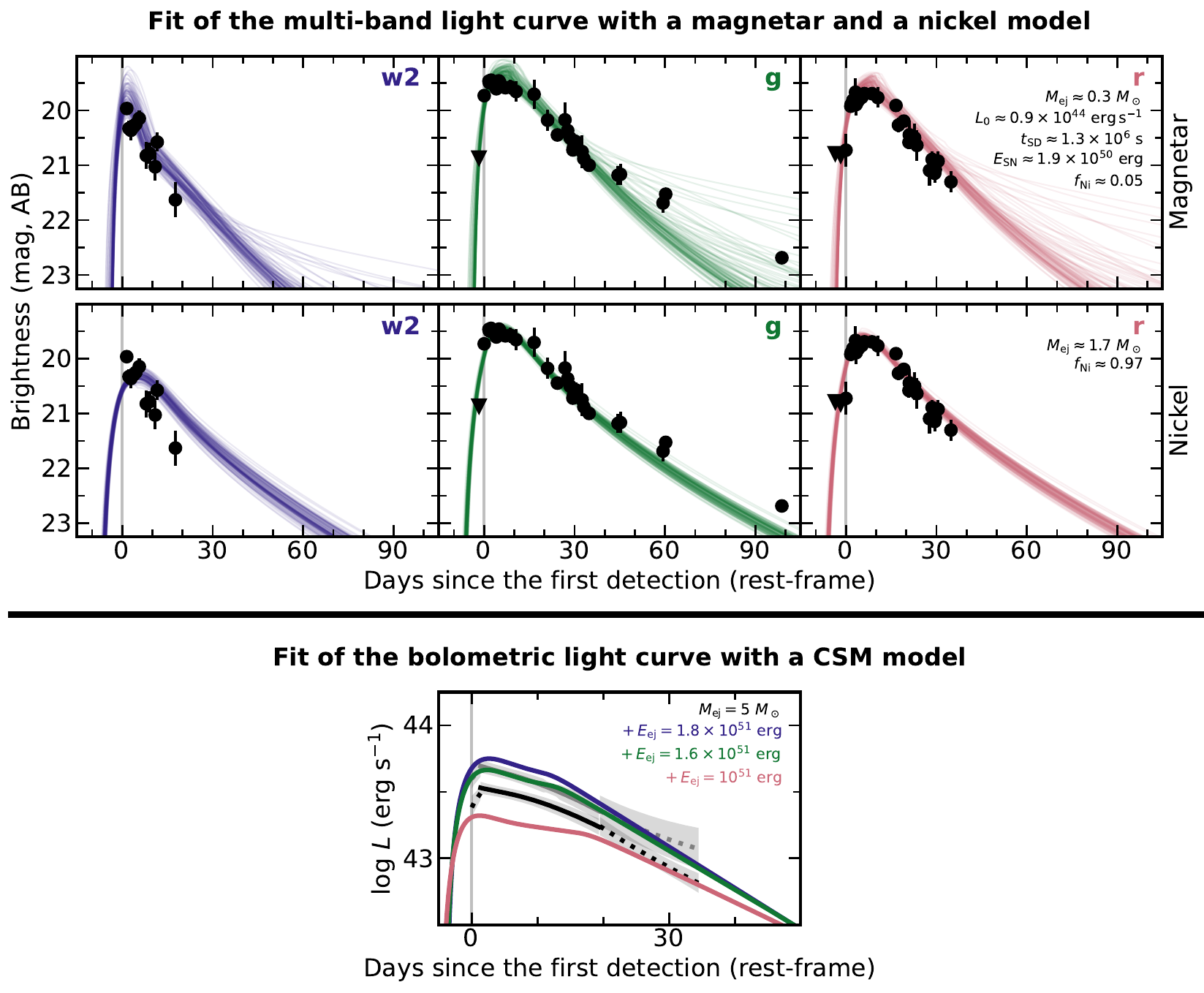}
    \caption{\textbf{Fits of the light curve of \sn with models of three different powering mechanisms.} The panels in the upper half show the results using magnetar and nickel models in the software package \program{Redback} \cite{Sarin2024a}. The bottom half shows a fit to the bolometric light curve using a CSM interaction model and the software package \program{CHIPS} \cite{Takei2022a, Takei2024a}. The CSM model can describe the observations. The mismatch between the observed and predicted rise can likely be mitigated with more complex CSM geometries and CSM density profiles than the ones considered here. The magnetar and nickel models can be excluded as the primary source of energy. The models do not simultaneously capture the rise, peak luminosity and peak time, and decline. Furthermore, the magnetar fit has an unphysically low opacity, and the nickel model requires an unphysically large nickel fraction. For illustration purposes, we only  show the results in three filters. Details about the modelling are provided in the \extmat \nameref{app:lc:modelling}. Non-detections are displayed as `$\blacktriangledown$'.
    }
    \label{fig:lc:redback}
\end{figure}

\begin{figure}[h!]
    \centering
    \captionsetup{name=\extmattwo Fig.}
    \includegraphics[width=\textwidth]{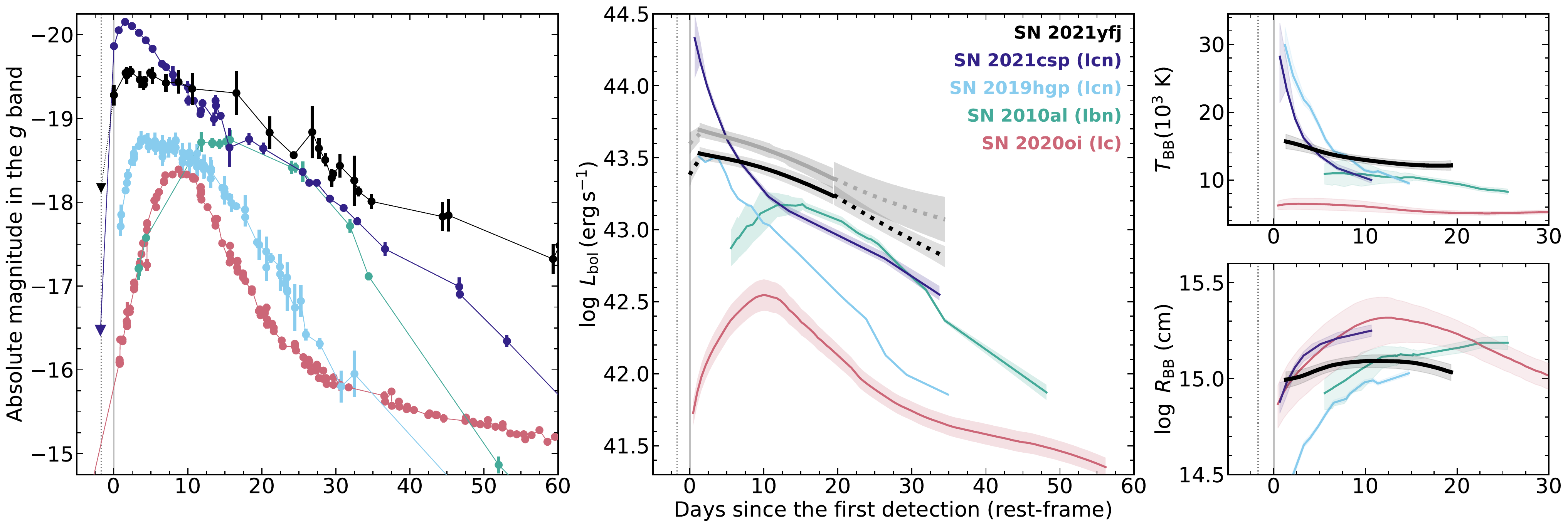}
    \caption{\textbf{Comparison of the light curves and blackbody properties of \sn with those of other interaction-powered SESNe and the Type Ic SN\,2020oi}. Compared to examples of interaction-powered SESN as well as the nickel-powered SN\,2020oi, \sn has a bright peak luminosity and a fast rise. Its blackbody radius and temperature evolve slowly in time compared to SNe~Icn. This gradual evolution is reminiscent of some SNe~IIn that are embedded in an optically thick CSM \cite[e.g.,][]{Soumagnac2020a}, whereas the rapid evolution of Type Icn events suggests a CSM that is significantly less optically thick (\method \nameref{app:lc:bolometrics}). The vertical dotted line in each panel indicates the date of the last nondetection of \sn. The statistical uncertainties at the $1\sigma$ confidence level are indicated as vertical error bars in the left panel and as bands in all other panels. Non-detections are displayed as `$\blacktriangledown$'.}
    \label{fig:lc:comparison}
\end{figure}

\begin{figure}[h!]
    \centering
    \captionsetup{name=\extmattwo Fig.}
    \includegraphics[width=0.41\textwidth]{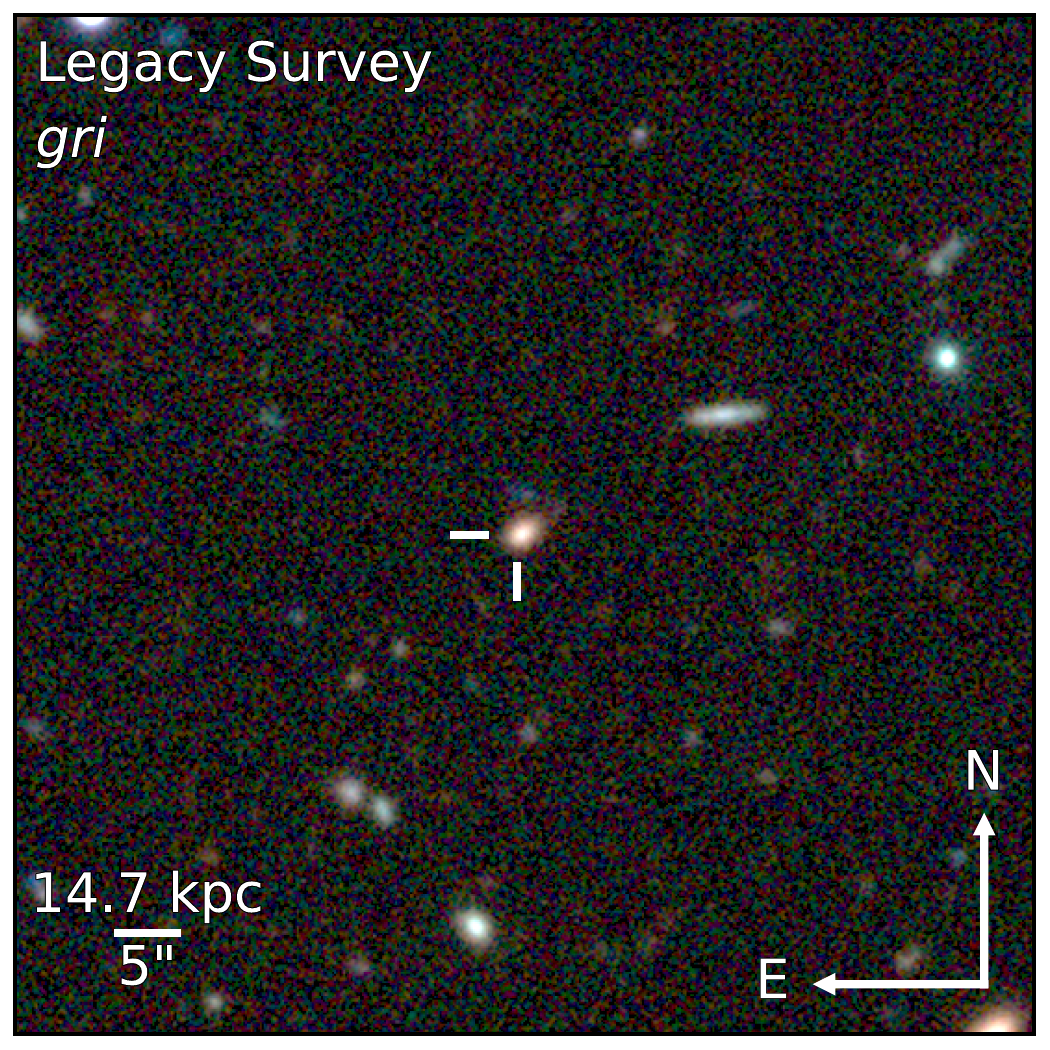}
    \includegraphics[width=0.58\textwidth]{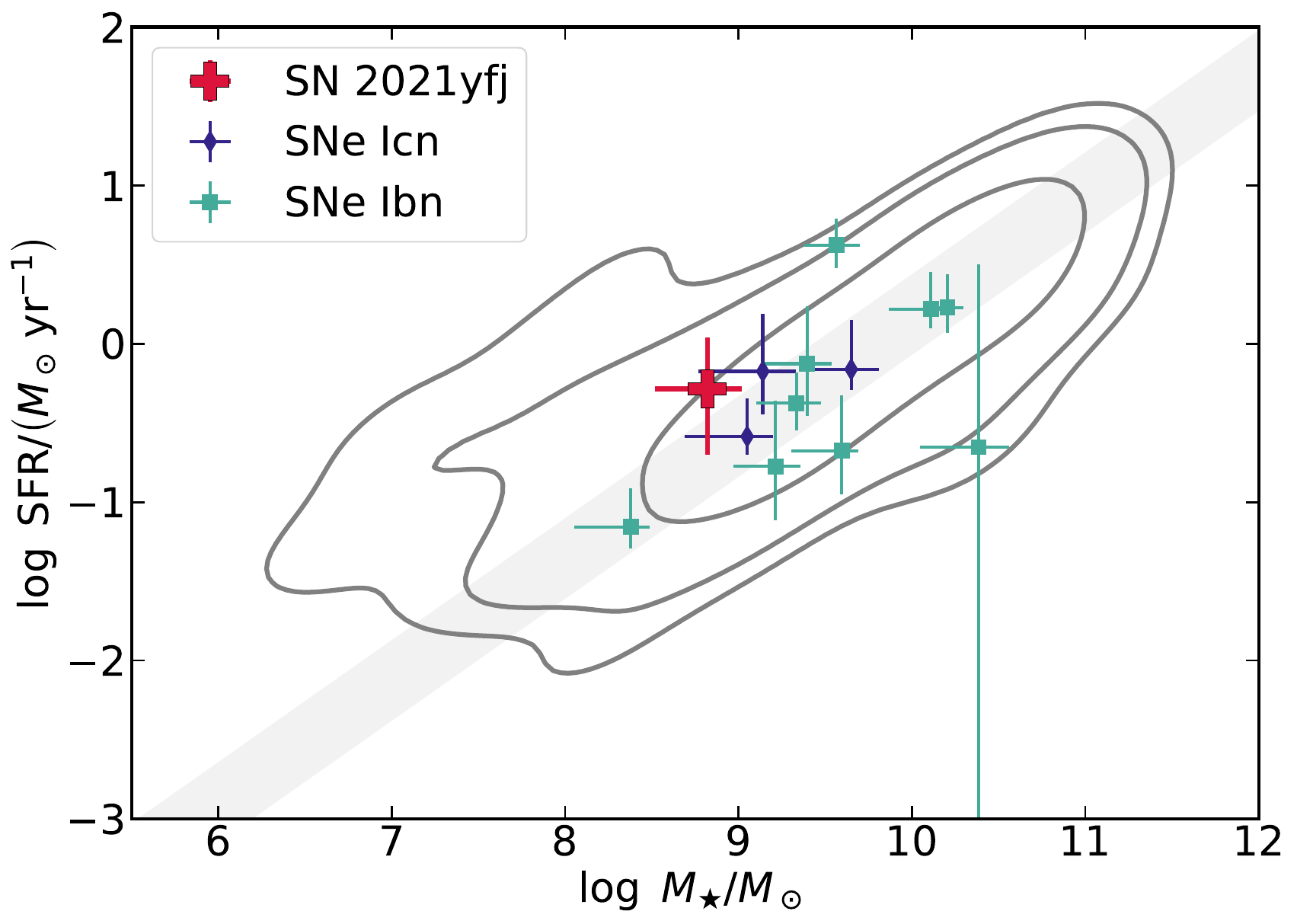}
    \caption{\textbf{The properties of \sn's host galaxy.}
    \textbf{\textit{Left:}}
    The SN position, marked by the crosshair, is located $\sim 1.2$~kpc south from the centre of its star-forming dwarf host galaxy ($M^{\rm host}_r \approx -18.1$~mag). 
    \textbf{\textit{Right:}}
    The host of \sn is a regular star-forming galaxy, demonstrated by its location with respect to the main sequence of star-forming galaxies (grey-shaded band, \cite{Elbaz2007a}). The properties are also consistent with hosts of core-collapse supernovae from the Palomar Transient Factory \cite{Schulze2021a} (grey contours indicate the region encircling 68, 90, and 95\%), including interaction-powered SESNe (Type Ibn and Icn SNe). The properties of all hosts were inferred from photometry using the software package \program{Prospector} \cite{Johnson2021a}.
    The statistical uncertainties at the $1\sigma$ confidence level are indicated. 
    }
    \label{fig:host:rgb}
    \label{fig:host:sed}
    \label{fig:host:mass_sfr}
\end{figure}

\clearpage
\input{tables/tab_snclassification}

%% file: tables/tab_snclassification.tex
\begin{table}
\captionsetup{name=\extmattwo Table}
\caption{The extension of the SN classification scheme after the discovery of \sn.
\label{tab:newsntype}}
\begin{tabular}{llll}
\toprule
Ejecta composition & CSM composition & SN Type   & SN Type\cite{Gal-Yam2017a}   \\
\midrule
H               & H       & SN IIn    & SN 0\,i0    \\
He              & He, (H) & SN Ibn    & SN 1\,i1    \\
C/O             & C/O     & SN Icn    & SN 2\,i2    \\
\rowcolor{mygray} O/Ne/Mg         & O/Ne/Mg & SN Idn    & SN 3\,i3    \\
\rowcolor{mygray} \textbf{O/Si/S}          & \textbf{O/Si/S}  & \textbf{SN Ien}    & \textbf{SN 4\,i4}    \\
\bottomrule
\end{tabular}
\footnotetext{
The SN classification scheme from Ref. \cite{Gal-Yam2017a} (last column) is a progression of the traditional system (second last column) 
The first number is the spectroscopic classifier of the ejecta composition: 0 = strong H features; 1 = strong He features but no H; 2 = strong C and O but no H and He; 3 = strong O, Ne, Mg, but no C, He, H; and 4 = strong Si, S, O but no Mg, Ne, He and H. The tag ``i'' stands for interaction followed by the composition of the material with which the SN interacts: 0 = strong H features, etc. 
The value of ejecta and CSM composition can take fractional values to indicate transitional objects.
The rows marked in grey are the new SN classes. \sn belongs to the hitherto unknown class of Type Ien SNe (bold).
}
\end{table}

%% file: methods.tex
\newpage

\section{Methods}
\subsection{Discovery}\label{app:discovery}

\sn, located at $\alpha=01^{\rm h}37^{\rm m}46\fs171$ and $\delta=-01\arcdeg15\arcmin17\farcs78$ [J2000.0], was discovered by the public ZTF Northern Sky Survey as ZTF21abzbmhz at 09:56 (UTC dates are used throughout this paper) on 7 September 2021 with an apparent magnitude of $r=20.82 \pm 0.30$~mag, about 1.7 rest-frame days after the last nondetection \cite{Munoz-Arancibia2021a}. The ZTF image-processing pipeline \cite{Masci2019a} generated an alert \cite{Patterson2019a} based on image subtraction \cite{Zackay2016a} with respect to a reference image. The alert was picked up by our custom ``infant supernovae" filter \cite{GalYam2019a, Bruch2021a} running on the ZTF Fritz Marshal system \cite{Walt2019a, Coughlin2023a}. It was identified by a duty astronomer, and follow-up observations were triggered using our standard methodology \cite{GalYam2011a, Bruch2021a}. The ALeRCE broker \cite{Foerster2021a} team independently discovered ZTF21abzbmhz in the ZTF alert stream. They were also the first to report ZTF21abzbmhz to the IAU Transient Name Server\footnote{\href{https://wis-tns.weizmann.ac.il/object/2021yfj}{https://wis-tns.weizmann.ac.il/object/2021yfj}} (TNS) \cite{Munoz-Arancibia2021a}. ZTF21abzbmhz was allocated the name \at on 7 September 2021. Later detections were reported by the Asteroid Terrestrial-impact Last Alert System (ATLAS; \cite{Tonry2011a}) survey on 12 September 2021 (internal name: ATLAS21bipz) and the Pan-STARRS Survey for Transients (PS; \cite{Huber2015a}) on 5 October 2021 (internal name: PS21ktg). On 3 September 2024, \at was designated the name \sn \cite{Gal-Yam2024c, Gal-Yam2024b}.
Unless stated otherwise, all times reported in this paper are with respect to the first detection and in the rest frame.

\subsection{Distance}\label{app:redshift}

The Keck spectrum from day 1 shows emission lines from hydrogen, oxygen and sulphur, produced by the ionised gas in the star-forming regions in the host galaxy, at a common redshift of $z=0.1386$. We refine redshift with the higher-resolution spectra obtained with X-shooter at the VLT. These spectra also show emission lines from hydrogen and oxygen. Averaging over all epochs, we measure a redshift of $z=0.13865\pm0.00004$. This is consistent with the redshift inferred from narrow absorption lines from \ion{Mg}{i}\,$\lambda$\,2852 and \ion{Mg}{ii}\,$\lambda\lambda$\,2796, 2803 from the host ISM, detected in the X-shooter spectra up to 6.1 days after discovery, and the redshift inferred from the host emission lines detected in the Keck spectra at days 128.1 and 132.5. We use the redshift of $z=0.13865$ as the SN redshift throughout the paper. The redshift translates to a luminosity distance of $676.4$~Mpc and a distance modulus of 39.15 mag using a flat $\Lambda$CDM cosmology with $H_0=67.7~\rm km\,s^{-1}\,Mpc^{-1}$, $\Omega_{\rm m}=0.31$, and $\Omega_{\Lambda}=0.69$ \cite{Planck2018a}, which we use throughout the paper.

\subsection{Pre-explosion Limits}

The ZTF survey started monitoring the SN field $\sim 3.2$~yr before its explosion. We search the archival ZTF data for pre-SN outbursts following the methods described by Ref. \cite{Strotjohann2021a}. We download IPAC difference images and compute the forced-photometry light curve at the SN position. After quality cuts, we are left with 612 pre-SN observations (74 points discarded) on 295 different nights in the ZTF $g$, $r$, and $i$ bands. We apply a baseline correction to ensure that the pre-SN light curve is centred around zero flux. The error bars are sufficiently large to account for the scatter of the pre-SN light curve and no upscaling of the statistical errors is required. We search for significant detections in unbinned and binned observations. Since the durations of the pre-explosion outbursts are unknown, we try seven different bin sizes between 1 and 90 days. We do not obtain $5\sigma$ detections for any of the searches. For week-long bins, the median limiting magnitude is $M > -18.8$~mag in  $g$ and $r$ and $M > -19.5$~mag in $i$. We can exclude outbursts that are brighter than those limits 50 (40) weeks before the explosion in $r$ ($g$). These limits are $\gtrsim1$~mag brighter than the most luminous precursors known \cite{Strotjohann2021a}, and hence do not pose meaningful limits on the outburst activity of \sn's progenitor shortly before the explosion.

\subsection{Light-Curve Properties}\label{app:lc}

The first detection is recorded at 09:56 on 7 September 2021, about 1.7 rest-frame days after the last nondetection. At the time of the discovery, \sn is very faint, $r=20.73\pm0.30$~mag [Milky Way (MW)-extinction corrected], but already luminous, $-18.3$~mag, owing to its large distance (\extmat \nameref{app:redshift}). It reaches its peak brightness in 2.3 and 5.8 rest-frame days in the $g$ and $r$ band, respectively. The MW-extinction corrected peak apparent magnitudes are $m_g\sim19.5\pm0.1$~mag and $m_r\sim19.7\pm0.1$ mag and translate to MW-extinction corrected, $K$-corrected \cite{Hogg2002a} absolute magnitudes of $M_g=-19.4\pm0.1$~mag and $M_r=-19.1\pm0.1$~mag. At the peak of the $g$-band light curve, \sn has a blue $g-r$ colour of $\sim-0.3$~mag, $K$-corrected and corrected for MW extinction. The high absolute magnitude at the time of discovery precludes estimating the explosion time as is commonly done for infant SNe (e.g., \cite{Bruch2021a, Bruch2023a, Miller2020a}). 

To put \sn in the context of other SN classes, we also compute the rise time from 50\% peak flux and how long the brightness stayed above 50\% of the peak flux in the $g$ band. We measure a rise time of 3 days and a duration of 21.7 days. Both measurements have an uncertainty on the order of a few days. For comparison, we choose an extended sample of extragalactic transients from the ongoing ZTF Bright Transient Survey. We apply the following selection criteria:
(i) no active galactic nuclei (AGNs),
(ii) the passing of the BTS data-quality cuts presented in Ref. \cite{Perley2020a},
(iii) well-sampled $g$ and $r$ light curves before, around, and after maximum brightness, 
(iv) a peak magnitude of $<19$ in $g$ and $r$ before MW-extinction correction,
(v) a well-measured rise and decline timescale in both bands,
(vi) a spectroscopic classification, and 
(vii) redshift information from either catalogues or the transient spectrum. In total, 4032 transients fulfil these criteria. The vast majority are Type Ia SNe (3178 objects, 79\%; e.g., \cite{Maguire2017a}) and Type II SNe (425 objects, 11\%; e.g., \cite{Arcavi2017a}). The remaining 10\% are other types of core-collapse supernovae (407 objects; e.g., \citenum{Filippenko1997a}, \citenum{Gal-Yam2017a}), tidal disruption events (16 objects; e.g., \cite{Gezari2021a}), intermediate luminosity red transients (3 objects; e.g., \cite{Bond2009a}), fast blue optical transients (1 object; e.g., \cite{Ho2023a}), gap transients (1 object; e.g., \cite{De2020a}) and luminous red novae (1 object; e.g., \cite{Pastorello2019a}). Among those objects, 22 objects are interaction-powered SESNe (Ibn: 20; Icn: 2). Their light curve properties (rise-time, duration, absolute peak magnitude and $g-r$ colour at peak) are computed following Ref. \cite{Perley2020a}. This comparison sample and \sn are shown in a 4-dimensional corner plot in the \extmat \figuretwo \ref{fig:lc:parameters}. Relative to the bulk population of SNe, the fast rise of \sn ($\sim$3\,days) represents its most extreme property. This, plus its relatively blue colour at peak ($g -r \approx -0.3\,\mathrm{mag}$) and moderately high luminosity ($M_g \approx -19.6\,\mathrm{mag}$) is largely consistent with the known population of SNe\,Ibn/Icn. Relative to the class of interacting SESNe, \sn, however, stands out for its long duration ($\sim$22\,days). In sum, aside from the short rise, \sn has a light curve that is largely consistent with the general properties of Type Ia SNe and regular core-collapse supernovae.

To quantify the peculiarity of the photometric evolution of \sn, we apply the Isolation Forest anomaly detection algorithm \cite{Liu2008} to determine whether \sn can be considered an outlier relative to other BTS sources. Briefly, an Isolation Forest builds a collection of decision trees and isolates individual sources by selecting a split point at random for a randomly selected feature within the feature space (4D as shown in the \extmat \figuretwo \ref{fig:lc:parameters}). Rare sources are, on average, isolated with fewer branch splittings within a tree than more common sources. Using the \program{scikit-learn} implementation of Isolation Forest \cite{Pedregosa11} with the default settings, we train the forest on the 4032 BTS sources. We then apply the forest to \sn and find that not only is it not an outlier, but also that $\approx15\%$ of the BTS sources are more ``rare'' than \sn in the 4D light-curve property feature space.  

\subsection{Bolometric Light Curve}\label{app:lc:bolometrics}

Following the procedure outlined in Refs. \cite{Nicholl2016a, Schulze2024a}, we compute the bolometric light curve over the wavelength range from 1,800 to 7,850~\AA\ (rest-frame), defined by the wavelength range of our photometric campaign from $w2$ to $z$ band. \figuretwo \ref{fig:lc:bolometric} shows the final bolometric light curve in black. A tabulated version is provided in the \suppmat \tabletwo \ref{tab:bolometrics}.

The solid black line in \figuretwo \ref{fig:lc:bolometric} shows the time interval of the bolometric light curve with the best spectral coverage. The blue $w2-r$ colour of $0.2\pm0.1$ mag at 1.5 days after the discovery of \sn points to a substantial contribution from the far UV. Linearly extrapolating the observed spectral energy distribution (SED) to shorter wavelengths yields a missing far-UV fraction of $39^{+10}_{-8}\%$ and $22^{+10}_{-7}\%$ at day 1.5 and day 19.7, respectively. The missing flux contribution beyond 1~$\mu$m is small. Fitting the observed SEDs from $w2$ to $z$ or from $u$ to $z$ yields a contribution of $\sim 5\%$ between 1 and 10~$\mu$m during the same time interval. The dark-grey curve in the \extmat \figuretwo \ref{fig:lc:bolometric} shows the bolometric light curve, including the two missing flux fractions. Other epochs have less well-observed SEDs, and we use the data between day 1.5 and day 19.7 to estimate bolometric corrections. The rising light curve was only observed in the $g$ and $r$ bands. This wavelength interval accounts for 15\% of the bolometric flux at day 1.5. Since SN ejecta cool with time, a constant bolometric correction will progressively underestimate the true bolometric flux toward earlier epochs. The fading light curve between days 19.7 and 32 was monitored from the $u$ to $i$ bands and in $gri$ between days 32.0 and 34.8. Similar to the data at the previous time interval, we compute bolometric corrections. The bolometric light curve of these time intervals is shown as dashed lines in the \extmat \figuretwo \ref{fig:lc:bolometric}. At $>34.8$~days after discovery, \sn is only detected in the $g$ band. We omit these data in the bolometric light curve.

Integrating the entire light curve yields a radiated energy of $0.6\times10^{50}~\rm erg$. Including the missing far-UV and IR contributions increases the radiated energy by a factor of $\sim 2$. Both values are comparable to those of other interaction-powered SESNe, such as SNe 2019hgp (Icn, $E_{\rm rad}\approx10^{50}~\rm erg$; \cite{Gal-Yam2022a}), 2020bqj (Ibn,  $E_{\rm rad}\approx10^{50}~\rm erg$; \cite{Kool2021a}), and 2021csp (Icn, $E_{\rm rad}\approx10^{50}~\rm erg$; \cite{Perley2022a}). 
Up to day 11.1, the spectra show a blackbody-like continuum with superimposed narrow emission lines but no broad metal absorption lines from the SN ejecta  (\suppmat \nameref{app:spec}). Fitting the photometry from 2,000 to 10,000~\AA\ (observer frame) with the Planck function yields a temperature of $\sim 16{,}000$~K at day 1.5 that gradually decreases to $\sim 12{,}000$~K in $\sim 3$ weeks and a radius that remains constant at $\sim 10^{15}~\rm cm$. The slow evolution of the bolometric light curve and the blackbody photosphere stand out compared to those of the well-observed interaction-powered SESNe 2010al, 2019hpg, and 2021csp (\extmat \figuretwo \ref{fig:lc:bolometric}). The bolometric flux of \sn decreases by less than 0.2~dex during the first two weeks since discovery, whereas the bolometric flux of the two Type Icn SNe faded by 1.0--1.2~dex and SN 2010al grew even brighter. The reasons for this are that \sn's photosphere remains at $\sim 10^{15}$~cm at all times and merely gradually cools. Such an evolution has been observed in SNe~IIn and was interpreted as the photosphere being located in the unshocked, optically thick CSM \cite{Ofek2014a, Soumagnac2020a}. In contrast to that, the photosphere of the Type Icn SNe 2019hgp and 2021csp undergoes a ballistic expansion, which points to a more optically thin CSM and, hence, a rapid decrease of the bolometric flux. 

\subsection{Light-Curve Modelling}\label{app:lc:modelling}

We model the multiband light curve with two distinct powering mechanisms using the open-source Bayesian inference software package \program{Redback} \cite{Sarin2024a}:
(i) radioactive decay of $^{56}$Ni \cite{Arnett1982a} and
(ii) spin-down of a rapidly spinning, highly magnetised neutron star (magnetar; \cite{Kasen2010a}) utilising the generalised magnetar model by Ref. \cite{Omand2024a}. 
The two models include a component to account for the loss of $\gamma$-ray trapping at late times \cite{Jeffery1999a}, which can increase the decline rate. Furthermore, we assume a Gaussian likelihood function, and we infer the model parameters with the \program{nessai} \citep{Williams2021a, Williams2021b, Williams2023a} sampler implemented in \program{Bilby} \cite{Ashton2019a, Romero-Shaw2020a}. The priors and marginalised posteriors of each model parameter are shown in the \suppmat \tabletwo \ref{tab:lc:redback}. 

The upper panels of the \extmat \figuretwo \ref{fig:lc:redback} show fits with the two magnetar and nickel models in three different bands. Both models are inadequate to describe the observations. The models are not able to fit the rise, peak and decline simultaneously. Furthermore, the magnetar model predicts an unphysically low opacity of $\kappa\approx0.01~\rm cm^2\,g^{-1}$, and the nickel requires an unphysically high nickel fraction of almost 100\%. Therefore, we reject these two powering mechanisms as the primary source of energy. 

Motivated by the lines of evidence for CSM interaction [\methods \nameref{app:spec} and \nameref{app:spec:comparison}], we model the bolometric light curve using the open-source code \program{CHIPS}\footnote{\href{https://github.com/DTsuna/CHIPS}{https://github.com/DTsuna/CHIPS}} \cite{Takei2022a, Takei2024a}. The code uses hydrodynamical calculations, together with radiative transfer, to calculate bolometric light curves powered by SN ejecta colliding with a dense CSM of an arbitrary density profile. We consider  homologously expanding ejecta with a density profile \cite{Matzner1999a} 
\begin{eqnarray}
\rho_{\rm ej} (r,t)
&=& \left\{ \begin{array}{ll}
t^{-3}\,\left[r/(gt)\right]^{-n} & (r/t > \upsilon_t),\\
t^{-3}\,(\upsilon_t/g)^{-n}\,\left[r/(t\upsilon_t)\right]^{-\delta}  & (r/t < \upsilon_t),
\end{array}\right.\nonumber
\label{eq:rho_ej}
\end{eqnarray}
where $n=10$ and $\delta=1$, commonly adopted for explosions for SESN progenitors. The constants $g$ and $\upsilon_t$ relate to the ejecta mass $M_{\rm ej}$ and energy $E_{\rm ej}$ as \cite{Moriya2013a}
\begin{eqnarray}
g = \left\{\frac{1}{4\pi\,(n-\delta)} \frac{[2\,(5-\delta)\,(n-5)\,E_{\rm ej}]^{(n-3)/2}}{[(3-\delta)\,(n-3)\,M_{\rm ej}]^{(n-5)/2}}\right\}^{1/n},~ \upsilon_t = \left[\frac{2\,(5-\delta)\,(n-5)\,E_{\rm ej}}{(3-\delta)\,(n-3)\,M_{\rm ej}}\right]^{1/2}.\nonumber
\end{eqnarray}
For the CSM density profile, we adopt a double power law, characterised by a shallow inner core and a steep drop generally found for simulations of envelope eruption \cite{Matzner1999a, Owocki2019a, Tsuna2021a},
\begin{equation}
   \rho_{\rm CSM}(r) = \hat{\rho}_{\rm CSM} \left[\frac{(r/r_{\rm CSM})^{n_{\rm in}/y}+(r/r_{\rm CSM})^{n_{\rm out}/y}}{2}\right]^{-y},\nonumber
\label{eq:rho_CSM}
\end{equation}
where $r_{\rm CSM}$ and $\hat{\rho}_{\rm CSM}$ set the radius and density at the transition of the two power laws, respectively. The remaining parameters $n_{\rm in}$ ($\approx 0$--$3$), $n_{\rm out}$ ($\approx 10$--$12$), and $y$ ($\approx 2$--$4.5$) are set by the envelope structure as well as the detailed hydrodynamics of the mass loss. However, the overall light-curve morphology is sensitive only to $n_{\rm in}$. We fix the other two as $n_{\rm out}=10$ (as adopted in the \method \nameref{app:spectral_modelling}) and $y=2$, inferred from simulations of partial envelope ejections from stripped progenitors \cite{Tsuna2023b}. \program{CHIPS} requires opacity tables for the CSM, which depend on its uncertain abundances. The spectra favour an O-dominant composition with enhanced Si/S/Ar and some He (\methods \nameref{app:spec}, \nameref{app:spec:helium}, and \nameref{app:spectral_modelling}). To reproduce this, we take the abundance of the surface of a stripped helium-poor stellar model with an initial mass of $29~M_\odot$ available in \program{CHIPS} \cite{Takei2024a}, and enhance the Si, S, and Ar mass fractions to those inferred from the \method \nameref{app:spectral_modelling} with carbon correspondingly reduced. The mass fractions adopted are 0.0081 (helium), 0.2041 (carbon), 0.6805 (oxygen), 0.0130 (neon), 0.0027 (magnesium), 0.05 (silicon), 0.03 (sulfur), 0.01 (argon), and small contributions of $<0.001$ for heavier metals. We use the Rosseland and Planck mean opacity tables for this composition, generated with \program{TOPS} \citep{Magee1995a}. While the adopted abundance may not exactly reflect that of the CSM, the bolometric light curves mostly depend on the dynamics of the interaction and much less on the composition details. We adopt a homologous CSM flow ($v\propto r$) as in the \method \nameref{app:spectral_modelling} and set the constant of proportionality to the CSM velocity of $2{,}000$~km\,s$^{-1}$ at $r=r_{\rm CSM}$, as observed in the early spectra (\method \nameref{app:spec}).

A successful fit to the bolometric light curve is shown in the bottom panel of \extmat \figuretwo \ref{fig:lc:redback}, with the fitting parameters shown in \suppmat \tabletwo \ref{tab:chips}. We find that in order to reproduce the bolometric light curve one needs 
(i) a moderately large explosion energy ($\sim 2\times 10^{51}$ erg) and CSM mass ($M_{\rm CSM}\gtrsim 1\,M_\odot$, with the lower limit being the estimated mass of the CSM swept up by the shock at day 30) for the long bright peak,
(ii) a shallow ($n_{\rm in}\approx 1$), extended ($r_{\rm CSM}\approx 5\times 10^{15}$ cm) inner CSM profile for the slow decline (see, e.g., \cite{Moriya2013a}), and 
(iii) a large $M_{\rm ej}$ ($\gtrsim M_{\rm CSM}$) so that the interaction power does not sharply decay owing to significant ejecta deceleration by the CSM.
The inferred high masses and energy for the ejecta and CSM are within the possible range for successive mass ejections in pulsational pair instability (PPI) models \cite{Fowler1964a, Barkat1967a, Rakavy1967a, Woosley2017a}. 
A pre-SN mass eruption in a lower-mass star cannot be ruled out, although a mechanism to eject such a large mass is less clear (\method \nameref{app:progenitor}).

\subsection{X-ray Emission}

The interaction of the SN ejecta with CSM and heating of the SN ejecta by a central engine (e.g., magnetar or black hole) can produce thermal X-ray emission \cite{Chevalier1992a, Chevalier1994a}. \sn was not detected in our \swift/XRT observations (\suppmat \tabletwo \ref{tab:xray}). The inferred upper limits between $10^{42}$ and a few $10^{43}~\rm erg\,s^{-1}$ (\suppmat \tabletwo \ref{tab:xray}) are comparable to the absorption-corrected luminosities of the X-ray brightest SNe \cite[][]{Chandra2012a, Dwarkadas2012a, Chandra2015a}, and thus do not place strong constraints on the lack of X-ray emission.

\subsection{Spectroscopic Evolution}\label{app:spec}

\extmat \figuretwo \ref{fig:spec:sequence} shows the spectral evolution from the rest-frame near-UV to near-IR from 1 to 50~days after the SN discovery. The spectra up to day 11 are characterised by a cooling blackbody (from $\sim22{,}000$ to $\sim15{,}000$~K; \figuretwo \ref{fig:lc:bolometric}) with superimposed narrow emission lines (width: $\sim 2{,}000~\rm km\,s^{-1}$, \extmat \figuretwo \ref{fig:spec:line_profiles}). Following Ref. \cite{GalYam2019b}, we use the atomic spectra database from the National Institute of Standards and Technology (NIST; \cite{NIST_ASD}) for the line identifications. We included all elements up to mass number 18 (argon) and created two ranked line lists for each ion sorted by
(i) the relative line intensity, and
(ii) the Einstein coefficient for spontaneous emission. We identify most lines as transitions from \ion{S}{iii}--\textsc{iv}, \ion{Si}{iii}--\textsc{iv}, and \ion{Ar}{iii}, and a very small minority of lines as transitions from \ion{Mg}{ii}, \ion{C}{iii}, and \ion{He}{i} (\figuretwo \ref{fig:spec:flash:id}, \extmat \figuretwo \ref{fig:spec:flash:id:full},  \suppmat \tabletwo \ref{tab:lines}). 

The \ion{Si}{iii}-\textsc{iv}, \ion{S}{iii}-\textsc{iv}, and \ion{Ar}{iii} are visible for $\lesssim6$ days (i.e., $\approx4$~days after the $g$-band maximum). As the bolometric luminosity (a proxy of the CSM interaction strength) and the blackbody temperature decrease, the highly-ionised species vanish, and the lines from singly ionised silicon and sulphur emerge. Between days 11 and 20, the spectrum transforms from a blackbody with narrow P~Cygni lines to a blue pseudocontinuum, akin to those of interaction-powered SESNe \cite{Hosseinzadeh2017a, Gal-Yam2022a, Perley2022a}, with superimposed intermediate (width: a few $1{,}000~\rm km\,s^{-1}$) emission lines from neutral and low-ionisation silicon, sulphur, helium, magnesium, and oxygen.

The emission features in the early spectra are visible as either pure emission lines or P~Cygni lines (examples shown in the \extmat \figuretwo \ref{fig:spec:line_profiles}). P~Cygni lines can be produced in the expanding SN ejecta, in an expanding shell of gas expelled by the progenitor before the explosion, or in a stellar wind. The absorption minima of the P~Cygni profiles are at $\sim 1{,}500~\rm km\,s^{-1}$ and the blue edge of the absorption component is at $\lesssim3{,}000~\rm km\,s^{-1}$ (best seen in the isolated \ion{Mg}{ii}\,$\lambda\lambda$\,2796, 2803 doublet). These velocities are significantly lower than those of typical SN ejecta (e.g., $\sim10{,}000~\rm km\,s^{-1}$; \cite{Liu2017a}). Instead, they are comparable to those of Wolf-Rayet stars \cite{Crowther2007a} and the CSM velocities of interaction-powered SESNe \cite{Gal-Yam2022a, Perley2022a}, meaning that \sn's ejecta interact with a fast-moving wind or shell of material that is rich in silicon and sulphur.

\subsection{Hydrogen and Helium Content}\label{app:spec:helium}

Hydrogen and helium constitute most of the baryonic matter in the Universe \cite{Asplund2009a}, and both elements play crucial roles in SN explosion and progenitor models \cite{Woosley2002a}. Hydrogen has strong features at 4{,}861 and 6{,}563 \AA\ \cite{NIST_ASD}. Throughout the evolution of \sn, we detect no hydrogen, neither in absorption nor emission. The blackbody temperature of \sn is similar to that of H-rich Type II SNe (e.g., \cite{Irani2024b}); therefore, \sn's progenitor must have lost its hydrogen envelope well before the SN explosion. \ion{He}{i} has its strongest optical transitions at 3{,}889, 4{,}471, 5{,}876, 6{,}678, and 7{,}065 \AA. The first high-resolution spectrum, obtained at 1.6~days after discovery, (\extmat \figuretwo \ref{fig:spec:flash:id:full}) shows well-defined emission lines at 3{,}889 and 5{,}876~\AA; the 7{,}065 \AA\ line is also detected but it is fainter than the 3{,}889 and 5{,}876 \AA\ lines. The intrinsically weak \ion{He}{i} line at 6{,}678 \AA\ is not detected, though it is redshifted to the Telluric A band (\extmat \figuretwo \ref{fig:spec:flash:id:full}). At the location of the 4,471 \AA\ line there is a broad emission complex that is likely due to multiple species, meaning we cannot reliably determine whether a \ion{He}{i} component exists at that wavelength. 

Using the method of Ref. \cite{GalYam2019b}, we find that 3{,}889 and 5{,}876 \AA\ are expected to be the strongest lines within the wavelength range we consider \cite{NIST_ASD}. A comparison with an early spectrum of the prototypical He-dominated Type Ibn SN 2006jc \cite{Anupama2009} indicates that this event 
shows only the 3{,}889 and 5{,}876~\AA\ lines at early times, further supporting the presence of \ion{He}{i} in the spectrum of \sn. The similarity between \sn and SN 2006jc is maintained during later phases (\figuretwo \ref{fig:spec:ibn_comparison}), where the broader \ion{He}{i} lines at 5{,}876~\AA\ and 7{,}065~\AA\ are most prominent in the spectra of both SNe \cite{Anupama2009,Pastorello2007a}. We also detect an emission line at 10,830~\AA\ at early and late times (\extmat \figuretwos \ref{fig:spec:flash:id:full}, \ref{fig:spec:sequence}), where \ion{He}{i} also has a strong transition. At late times, the 5{,}876-\AA\ feature of \sn reveals time-variable shoulders due to the contribution from singly-ionised silicon and sulphur (\extmat \figuretwo \ref{fig:spec:line_profiles}).

Therefore, we conclude that the spectroscopic data provide evidence for the existence of helium within the emitting material throughout the SN evolution.

\subsection{Comparison With Interaction-powered SESNe}\label{app:spec:comparison}

To understand the peculiarity of \sn, we compare its spectral evolution to that of other interaction-powered SESNe. We chose SNe 2006jc and 2010al (as archetypes of the Type Ibn class; \cite{Foley2007a, Pastorello2007a, Pastorello2008a}), and SNe 2019hgp and 2021csp (as archetypes of the Type Icn class; \cite{Gal-Yam2022a, Perley2022a}). SNe 2019hgp and 2021csp were detected within $<2$~days of their explosion dates, and spectra were acquired within hours after their discovery, offering an excellent opportunity to search for silicon, sulphur, and argon in their earliest spectra.

\figuretwo \ref{fig:spec:ibn_comparison} shows snapshots of the spectral evolution before peak light (top), at maximum  (centre), and more than one month after peak (bottom). The different SN classes evolve similarly. Up to the time of maximum light, the spectra are characterised by a thermal spectrum cooling from a few 10,000~K to 10,000--15,000~K and a series of narrow emission lines [full width at half-maximum intensity (FWHM) $<1000~\rm km\,s^{-1}$]. Well after maximum brightness, a blue pseudocontinuum develops, produced by a forest of iron emission lines in the CSM, and a small number of emission lines with a FWHM of several $1000~\rm km\,s^{-1}$. Depending on the elemental composition of the CSM, different emission lines are visible: Type Ibn -- helium (primarily), hydrogen, carbon, and oxygen; Type Icn -- carbon, neon, oxygen (primarily) and helium; and \sn -- silicon, sulphur, argon (primarily) and helium. Furthermore, SNe 2010al and 2021csp show conspicuous emission from \ion{Ca}{ii} at 8,500~\AA. This feature is absent in \sn, which is puzzling considering that argon, calcium, silicon, and sulphur are the ashes of the oxygen-burning phase (e.g., \cite{Woosley2005a}).

In conclusion, the nondetection of silicon, sulphur, and argon in Type Ibn and Icn SNe and likewise the nondetection of hydrogen, nitrogen and oxygen and the weak presence of carbon and helium in \sn, is not due to differences in the ionising radiation field, insufficient data quality, or the wavelength coverage. Instead, it reflects differences in the elemental composition of the CSM and, therefore, in the progenitor stars (e.g., \cite[][]{Dessart2022a}). 

\subsection{A New Type of Supernova}\label{app:newsntype}

The classification of SNe is fundamentally based on spectroscopy \cite{Filippenko1997a, Gal-Yam2017a}. The type of a SN is determined by the dominant spectroscopic features in its peak-light spectra: Type II SNe are dominated by hydrogen lines, Type Ib SNe by helium, and Type Ic SNe lack both hydrogen and helium. This sequence is assumed to reflect the amount of stripping of the progenitor stars, with those of  SNe~II having retained much of their initial hydrogen envelope, those of SNe~Ib having lost the hydrogen layer but retaining the He-rich layer below, and those of SNe~Ic having lost all or most of the He-rich layer. Adding the suffix ``n'' to the SN type is used to indicate the presence of relatively narrow spectral lines that arise from the SN progenitor having been surrounded by slowly-moving CSM whose composition reflects that of the outer stellar layer at the time of explosion or shortly prior. Thus, SNe~IIn  are surrounded by H-rich CSM, SNe~Ibn have He-rich CSM, and SNe~Icn have C/O-rich CSM. Following the theoretical shell structure of massive stars, one would expect that even further stripping would lead to the formation of stars whose outer layers are dominated by O/Ne/Mg and later O/Si/S, with natural designations of Type Id and Ie; events with narrow CSM lines would then be denoted by Idn and Ien (\extmat \tabletwo \ref{tab:newsntype}). Our observations, presented in the \suppmat \nameref{app:spec}, \nameref{app:spec:helium}, and \nameref{app:spec:comparison}, suggest that \sn is indeed the first example of a Type Ien SN \cite{Gal-Yam2024a}. This discovery also implies the existence of Type Id, Idn and Ie SNe. A late-time spectrum of the Type Ic SN\,2021ocs \cite{Kuncarayakti2022a} may indicate that its progenitor was more rich in O/Mg than the progenitors of other Type Ic SNe. SN\,2021ocs could be Id or Ic/Id transitional SN. However, the evidence comes solely from a nebular spectrum, whereas SN classifications are based on their peak-light spectra. While it is too early to claim the detection of a Type Id SN, the discovery of SN\,2021ocs is very intriguing.

\section{Spectral Modelling}\label{app:spectral_modelling}

Using the approach of Ref. \cite{Dessart2022a}, we simulate the early-time spectra of \sn by adopting ejecta with a density profile of $\rho \propto r^{-10}$, a velocity of $1000~\rm km\,s^{-1}$ at $10^{15}~\rm cm$ (together with a homologous flow, i.e., $r/v = \rm age = 116~days$), and a composition with the following mass fractions: 0.786 (oxygen), 0.1 (neon), 0.05 (silicon), 0.03 (sulphur), 0.01 (argon), 0.01 (magnesium), 0.001 (calcium), and solar abundance for iron, cobalt, and nickel. This composition is representative of the O/Si shell in a massive He-star model at the time of explosion, such as the he12 model used in Ref. \cite{Dessart2021a}. The density is scaled so that the total Rosseland-mean optical depth of the ejecta is 40, assuming a mean opacity of 0.1 $\rm cm^2\,g^{-1}$, roughly comparable with the results from the radiative-transfer calculation. This yields a total mass of $3.24~M_\odot$, which is quite substantial. Finally, a power of $3\times10^{43}~\rm erg\,s^{-1}$ is injected into the inner regions of these ejecta at $v_{\rm deposition}=1000~\rm km\,s^{-1}$ over a characteristic scale of ${\rm d}v=200~\rm km\,s^{-1}$. The deposition profile goes as $\exp [- \left(v-v_{\rm deposition}\right)^2\,/\,{\rm d}v^2 ]$, and the volume-integrated power is normalised to $3\times10^{43}~\rm erg\,s^{-1}$. Given these initial conditions, the radiative-transfer solution computes the temperature, ionisation, etc. We then compute steady-state, non-local thermodynamic equilibrium (NLTE) radiative-transfer models \cite{Dessart2022a}.

The results for the UV-optical spectrum are shown in \figuretwo \ref{fig:spec:model}. The model spectra contain numerous lines of silicon and sulphur despite the relatively low abundance of a few 0.01, revealing that these elements have a strong absorption power, like iron, even for a modest abundance. Additional tests, in which we raise the silicon and sulphur abundances until they reach values of 0.3--0.5, yield somewhat stronger silicon and sulphur lines, although not so much stronger, but progressively in tension with the observations. Decreasing the mass fraction below 1\% gives weaker features, again in tension with observations. Hence, it seems that a mass fraction of a few 0.01 of silicon and sulphur is sufficient to explain the optical spectra of \sn. These explorations remain somewhat short of the true complexity of \sn. Indeed, no O-rich or Si-rich material in a massive star is also rich in helium, which is in tension with the presence of \ion{He}{i} lines detected in \sn (\method \nameref{app:spec:helium}). One way to accommodate this peculiarity is by invoking an asymmetric configuration in which He-rich material would be present in some ``equatorial region" and O/Si-rich material interacts with that material, producing an interacting SN, with emission coming both from that He-rich CSM and the Si/O-rich ejecta. Further work is needed to investigate this aspect thoroughly. 

\subsection{Host Galaxy}\label{app:host}

\sn's host galaxy was detected in several broad-band filters from the rest-frame UV to near-IR ($m_r \approx 21$~mag; \suppmat \tabletwo \ref{tab:phot:host}). The left panel of the \extmat \figuretwo \ref{fig:host:rgb} shows a false-colour image of the host galaxy, built with $gri$ images from the DESI Legacy Imaging Surveys \cite{Dey2018a} and the software package \program{STIFF} \cite{Bertin2012a} version 2.4.0. The SN explosion site is $\sim 1.2$~kpc south of the galaxy centre. To infer the mass and star-formation rate of the host, we model the observed SED (\suppmat \tabletwo \ref{tab:phot:host}) with the software package \program{Prospector} \cite{Johnson2021a} version 1.1.\footnote{\program{Prospector} uses the \program{Flexible Stellar Population Synthesis} (\program{FSPS}) code \cite{Conroy2009a} to generate the underlying physical model and \program{python-fsps} \cite{ForemanMackey2014a} to interface with \program{FSPS} in \program{python}. The \program{FSPS} code also accounts for the contribution from the diffuse gas based on the \program{Cloudy} models from \cite{Byler2017a}. We use the dynamic nested sampling package \program{dynesty} \cite{Speagle2020a} to sample the posterior probability.} We assume a Chabrier initial-mass function (IMF; \cite{Chabrier2003a}) and approximate the star-formation history (SFH) by a linearly increasing SFH at early times followed by an exponential decline at late times [functional form $t \times \exp\left(-t/t_{1/e}\right)$, where $t$ is the age of the SFH episode and $t_{1/e}$ is the $e$-folding timescale]. To account for any reddening between the expected and the observed SED, we use the Calzetti attenuation model \cite{Calzetti2000a}. The priors of the model parameters are set identical to those used by Ref.~\cite{Schulze2021a}. The host galaxy has a stellar mass $\log\,M_\star/M_\odot=8.9\pm0.2$, a star-formation rate of $0.07^{+0.10}_{-0.02}~M_\odot\,{\rm yr}^{-1}$, an age of $4^{+4}_{-2}$~Gyr, and an attenuation of the stellar component of $E(B-V)=0.05^{+0.07}_{-0.03}$~mag ($\chi^2/\rm n.o.f. = 18.25 / 11$, where n.o.f. is the number of photometric filters used in the SED modelling). The star-formation rate is comparable to typical star-forming galaxies of that stellar mass (grey band in \extmat \figuretwo \ref{fig:host:mass_sfr}; \cite{Elbaz2007a}). The mass and star-formation rate are also similar to the SN host galaxies from the Palomar Transient Factory (grey contours; \cite{Law2009a, Rau2009a, Kulkarni2013a, Schulze2018a}), including interaction-powered SESNe (colour-coded; the values of the SNe Icn were taken from Refs. \cite{Gal-Yam2022a, Perley2022a, Pellegrino2022a}).

The SN spectra reveal emission lines from the ionised gas in \ion{H}{ii} regions along the line of sight, summarised in the \suppmat \tabletwo \ref{tab:host_ism}. Their luminosities and flux ratios allow us to determine the metallicity of the gas in the star-forming regions, the metal enrichment, and the level of attenuation. The MW-extinction corrected H$\gamma$/H$\beta$ and H$\delta$/H$\beta$ flux ratios are $0.46\pm0.02$ and $0.23\pm0.02$, respectively. Both values are consistent with the theoretically predicted values of 0.47 and 0.26, assuming typical conditions of star-forming regions: Case B recombination, electron temperature of $10^4~\rm K$, and electron density of $10^2~\rm cm^{-3}$ \cite{Osterbrock2006a}. The nominal excess in the flux ratio translates to 
$E_{\rm host}(B-V) = 0.07\pm0.06$~mag, assuming the Calzetti attenuation model with $R_V=4.05$. Owing to the small amount of attenuation and its large statistical measurement error, we assume negligible extinction for all SN properties. The H$\alpha$ luminosity and the level of star formation are tightly correlated \cite{Kennicutt1998a}. The attenuation-corrected star-formation rate is $0.17 \pm 0.03~M_\odot\,{\rm yr}^{-1}$ using Ref. \cite{Kennicutt1998a} and Ref. \cite{Madau2014a} to convert from the Salpeter IMF (assumed in Ref. \citenum{Kennicutt1998a}) to the Chabrier IMF (assumed in our galaxy SED modelling). 
Both the attenuation and the star-formation rate are consistent with the values derived from the SED modelling. The metallicity of the star-forming region can be determined from the ratios between H$\alpha$, H$\beta$, [\ion{N}{ii}]\,$\lambda$\,6584, and [\ion{O}{iii}]\,$\lambda$\,5007 \cite{Pettini2004a}. Using this O3N2 diagnostic together with the parameterisation from Ref. \cite{Curti2017a}, we infer a gas-phase metallicity of $0.53\pm0.01$ solar, a normal value for a galaxy of that mass \cite{Andrews2013a}. 

We note that \sn exploded close to the centre of its host galaxy, and the slit covered a large fraction of the host galaxy. The properties reported above are, hence, representative of the SN explosion site and the entire galaxy.

\section{Event Rate}\label{app:rates}

The WISeREP archive contains $>$ 400 public spectra for $> 70$ SNe Ibn/Icn. We examined all spectra to determine whether any of the objects showed narrow silicon or sulphur lines. None of the objects has spectra exhibiting silicon, sulphur, and argon lines. This reveals that \sn is the first member of a previously unknown supernova class and that Type Ien SNe are an even rarer class of objects than the already rare Type Ibn and Icn SNe (SNe Ibn, 0.1--0.5\%; SNe Icn, 0.005--0.05\% of the total CCSN rate; \cite[][]{Perley2022a}).

The nondetection of Type Ien SNe in the ZTF Bright Transient Survey allows us to place an upper limit on their volumetric rate. Figure 9 in Ref \cite{Perley2020a} shows the relationship between the volumetric rate as a function of peak absolute magnitude derived from the ZTF Bright Transient Survey. Using the same methodology together with the 6-year BTS sample and assuming that SNe Ien reach an absolute magnitude of $<-19$~mag at peak, their volumetric rate is $<30~\rm Gpc^{-1}\,yr^{-1}$ at $95\%$ confidence (Poisson statistics), which is roughly $<1/1{,}000$ the rate of SNe\,Ib/c \cite{Li2011a} and $<1/3{,}000$ of the total CCSN rate \cite{Perley2020a}.

\section{Progenitor Scenarios}\label{app:progenitor}

Knowing that a moderate amount of silicon and sulphur suffices to explain the features of the early-time spectra of \sn, we now explore possible progenitor channels.

\subsection{A High-mass Massive Star}

Massive stars can lose a substantial amount of their birth mass through stellar winds \cite{Vink2022a, Smith2014a}, eruptions \cite{Humphreys1994a, Smith2014a}, and interaction with a companion star \cite{Podsiadlowski1992a, Marchant2023a}. First, we focus on stars that have lost their entire hydrogen envelope, so-called He stars, with a pre-supernova mass between 30 and 133~$M_\odot$. During the oxygen-burning phase of such a star, $e^+e^-$ pairs are formed, reducing the radiation pressure that supports the star against gravitational collapse (so-called pair instability; \cite{Fowler1964a, Barkat1967a, Rakavy1967a}). As a result, implosive oxygen burning can produce enough energy to unbind a substantial amount of the stellar envelope. He cores above $\gtrsim64~M_\odot$ experience a single violent pulse that unbinds the entire star \cite{Heger2002a, Umeda2002a, Kasen2011a, Kozyreva2017a, Gilmer2017a}. Less-massive stars can encounter pair instability a few times and eject shells of increasingly metal-rich CSM. The collisions of shells can produce luminous optical transients, so-called pulsational pair-instability SNe (PPISNe; \cite[][]{Woosley2007a, Yoshida2016a, Woosley2017a}). The specific mass limits depend on the uncertain rate of the $^{12}$C($\alpha,\gamma)^{16}$O reaction rate \cite{Farmer2020a, Woosley2021a, Farag2022a}, but for reasonable choices, a qualitatively similar behaviour is observed in the transition region between PPISNe and PISNe, wherever it occurs. PPISN models usually do not produce shells that contain much  silicon and sulphur. Usually, the outer layers of helium, carbon, oxygen, magnesium, and neon are ejected. In the transition between PPISN and PISN, the first pulse can eject an arbitrary amount of mass and expose the oxygen convective shell. This extensive convective shell encompasses much of the star and may be mildly enhanced in newly synthesised silicon.

Among the available PPISN models, the He60/61 model of Ref. \cite{Woosley2017a} with a helium-core mass of 60--61~$M_\odot$ is of particular interest. It has an oxygen convective shell with a composition similar to that of the he12 model whose predicted spectral features matched the observed ones at early times (\method \nameref{app:spectral_modelling}). In the He61 model, the first pulse ejects $19~M_\odot$, exposing the O/Ne shell. The remaining $42~M_\odot$ are almost unbound, but eventually, the star contracts and reencounters pair instability. Enough nuclear fuel is left for a few final pulses that happen in rapid succession before the iron core directly collapses into a black hole. With each new pulse, a new shell (moving with a velocity of a few $1{,}000~\rm km\,s^{-1}$) is enriched in increasingly heavier elements, and, eventually, material from the oxygen convective shell, which could be enhanced in silicon, can be expelled. The collisions between the shells (pulse 2 and later) can produce light curves with rise times as short as a few days, durations of several tens of days, and peak luminosities reaching a few $10^{43}~\rm erg\,s^{-1}$. Collisions between the shells can radiate up to $5\times10^{50}~\rm erg$ and produce spectra that are dominated by interaction \cite{Woosley2017a}. However, the exact properties are more uncertain. They are highly sensitive to the properties of the first pulse, the kinematics of the shells, the mass they carry, the time between the pulses, mixing processes, and details of the simulations \cite{Chen2014a, Woosley2017a, Leung2019, Chen2023c}.

In the PPISN scenario, \sn could be the product of a collision between the last shells ejected before the progenitor star collapsed to form a black hole. Qualitatively, the PPI model explains many of the observed properties, such as the wind velocity of $3{,}000~\rm km\,s^{-1}$, presence of silicon, sulphur, and argon in the early spectra, interaction-dominated spectra throughout the entire evolution (\method \nameref{app:spec}), the rise time of 3 days, the peak luminosity of (3--5) $\times10^{43}~\rm erg\,s^{-1}$, and the radiated energy of (0.5--1) $ \times10^{50}~\rm erg$ (\method  \nameref{app:lc}). It could also provide a mechanism to strip a star to its oxygen convective shell. The host metallicity of $\lesssim0.5$ solar (\method \nameref{app:host}) and low event rate of \sn-like transients (\method \nameref{app:rates}) match further predictions of PPISN models. 

The presence of helium (\method \nameref{app:spec:helium}) in the spectra of \sn is puzzling. There was helium on the surface of the 61/62-$M_\odot$ models, but it was ejected about a thousand years before the explosion and now resides at  $\sim 10^{18}$~cm, maybe less if it collides with dense CSM. Since massive stars tend to live in binary systems, it may not be too unlikely to have a helium-star companion with a strong wind. All stars this massive have nearly the same lifetimes burning hydrogen and helium ($3\times10^6$ and $3\times10^5$ yr, respectively), so a coeval star could be in a similar stage of evolution.

\subsection{A Low-mass Massive Star}

In the regime of less-massive He stars, a few scenarios could give rise to \sn, as follows.
\newline
\newline
\noindent\textbf{Scenario A} --- 
Helium stars with pre-supernova masses in the range 2.0 to $2.6~M_\odot$ have a complex final evolution that is strongly influenced by electron degeneracy \cite{Woosley2015a, Woosley2019a}. To make a low-mass helium star, a close interacting binary is a necessary starting point. Both oxygen and silicon burning ignite far off centre in these stars, and the fusion propagates inward as ``flames'' bounding a molecular-weight inversion that might be unstable \cite{Woosley2015a}. The stars also have unusually large radii ($\sim10^{13}$ cm) that they develop after core helium depletion. Some also experience strong, degenerate silicon flashes that, while energetically incapable of exploding the entire star, can eject all or part of the matter outside the silicon core. A relevant scenario would be
(i) the silicon flash plus any residual binary interaction removes most of the matter external to the oxygen-burning shell;
(ii) during the following months, the inwardly propagating flame powers a strong wind that ejects more silicon-rich matter; and
(iii) a terminal iron-core collapse creates an outgoing shock with $\sim10^{50}$ erg.
The bright supernova results from the terminal explosion interacting with the wind and ejected shell, as shown in Figure 14 of Ref. \cite{Woosley2019a}. This could be a more common occurrence than a PPISN and does not require low metallicity or a high star-formation rate. Helium would also be present in the recently ejected matter. The difficulty is the many uncertainties surrounding the energy and timing of the silicon flash; the propagation, in more than one dimension (1D), of the flames; and the extent to which the flash and winds uncover silicon-rich material prior to the terminal explosion. We note that based on simulations by Ref. \cite{Dessart2022a}, the CSM around such low-mass stars is expected to consist mostly of helium, which leads to strong helium features throughout the entire evolution, inconsistent with observations of \sn [\extmat \figuretwo \ref{fig:spec:flash:id:full}, \ref{fig:spec:sequence}]. This channel might be ideal for Type Ibn SNe \cite{Dessart2022a}, but less so for \sn.
\newline
\newline
\textbf{Scenario B} --- Some massive stars produce jets during their terminal explosion \cite{Woosley2006a, Hjorth2012a}. The appearance of Si/S material in the outer layer could be the result of a jetted explosion where jets lift and drag material from the stellar interior onto the outer layers. In this scenario, \sn would have to be observed from a preferred direction close to on-axis. This might also produce a $\gamma$-ray flash as seen in long-duration GRBs, which are connected with the explosion of very massive stars. We found no $\gamma$-ray flash or an afterglow in the X-ray or the optical cospatial with \sn within 2 days before the discovery of \sn (\suppmat \nameref{app:obs:xray}).

Considering \sn's moderately high redshift, an intrinsically weak $\gamma$-ray flash would likely have evaded detection with current $\gamma$-ray satellites \cite{Pian2006a, Starling2011a}. Furthermore, a baryon loading of as little as $10^{-4}~M_\odot$ is expected to stifle the formation of a relativistic jet \cite{Piran2004a}, leading instead to the formation of a nonrelativistic outflow. An outflow should not only dredge up silicon and sulphur; it should also transport carbon, oxygen, and neon from the layers  between the inner Si/S shell and the other He shell. However, strong lines from carbon, oxygen, and neon are absent in the spectra of \sn. 

\subsection{A Merger of Two Compact Objects}

In the following, we explore whether silicon and sulphur could be formed on the surface of a white dwarf or possibly a neutron star. Simulations by Ref. \cite{Khokhlov1986a} showed that helium burning at low densities ($\sim10^9~\rm g\,cm^{-3}$) and low temperatures ($\sim10^9~\rm K$) produces silicon and sulphur. Helium burning is an exothermal process, though. The temperature increase would cease the production silicon and sulphur and instead continue fusing the ashes to nickel \cite{Khokhlov1986a}. The excess energy could be dissipated through the expansion of the gas. However, this would also skew the production toward heavier nuclei \cite{Khokhlov1986a}. Cooling by emitting neutrinos is not possible since silicon and sulphur are stable against $\beta$ decay in these conditions. Simulations of helium burning on the surface of C/O white dwarfs \cite{Waldman2011a} revealed that the density on their surface is always $\ll10^9~\rm g\,cm^{-3}$, making the production of silicon and sulphur subdominant to the more common elements, such as calcium, magnesium, and iron-group elements. We note, though, that helium burning on the surface of different types of compact objects (white dwarfs and neutron stars) is underexplored, and detailed simulations are needed.

\input{acknowledgements}
\input{contributions}

\section{Data Availability}

The reduced spectra and photometry of \sn will be made available via the WISeREP archive and the journal webpage after the acceptance of the paper. Data from ESO, Keck, NOT, \swift, and ZTF can be obtained from their designated public data repositories.

\section{Code Availability}
Much analysis for this paper has been undertaken with publicly available codes. The details required to reproduce the analysis are contained within the manuscript. 

%% file: acknowledgements.tex
\section{Acknowledgements}

M.~W. Coughlin acknowledges support from the U.S. National Science Foundation (NSF) with grants PHY-2308862 and PHY-2117997. 
A. V. Filippenko's group at UC Berkeley is grateful for financial assistance from the Christopher R. Redlich Fund, Gary and Cynthia Bengier, Clark and Sharon Winslow, Alan Eustace (W.Z. is a Bengier-Winslow-Eustace Specialist in Astronomy), William Draper, Timothy and Melissa Draper, Briggs and Kathleen Wood, Sanford Robertson (T.G.B. is a Draper-Wood-Robertson Specialist in Astronomy), and many other donors.   
A. Gal-Yam's research is supported by the ISF GW excellence centre, an IMOS space infrastructure grant and BSF/Transformative and GIF grants, as well as the Andr\'e Deloro Institute for Space and Optics Research, the Center for Experimental Physics, a WIS-MIT Sagol grant, the Norman E Alexander Family M Foundation ULTRASAT Data Center Fund, and Yeda-Sela; A. Gal-Yam is the incumbent of the Arlyn Imberman Professorial Chair.
N. Kne\v zevi\'c was supported by the Ministry of Science, Technological Development and Innovation of the Republic of Serbia (MST-DIRS) through contract no. 451-03-66/2024-03/200002 made with the Astronomical Observatory (Belgrade) and contract no. 451-03-66/2024-03/200104 made with the Faculty of Mathematics at the University of Belgrade.
R. Lunnan acknowledges support from the European Research Council (ERC) under the European Union's Horizon Europe research and innovation programme (grant agreement 1010422). 
K. Maeda acknowledges support from the JSPS KAKENHI grant JP20H00174 and JP24H01810. 
A. A. Miller and S. Schulze are partially supported by LBNL Subcontract 7707915.  
N. Sarin acknowledges support from the Knut and Alice Wallenberg foundation through the ``Gravity Meets Light" project. 
D. Tsuna is supported by the Sherman Fairchild Postdoctoral Fellowship at the California Institute of Technology. 
Y. Yang appreciates the generous financial support provided to the supernova group at U.C. Berkeley by Gary and Cynthia Bengier, Clark and Sharon Winslow, Sanford Robertson, and numerous other donors.
We appreciate the excellent assistance of the staff at the various observatories where data were obtained.
U.C. Berkeley undergraduate student Evelyn Liu is thanked for her effort in taking Lick/Nickel data. 

Based in part on observations obtained with the 48-inch Samuel Oschin Telescope and the 60-inch Telescope (P60) at the Palomar Observatory as part of the Zwicky Transient Facility project. ZTF is supported by the U.S. NSF under grants AST-1440341 and AST-2034437, and a collaboration including current partners Caltech, IPAC, the Weizmann Institute of Science, the Oskar Klein Centre at Stockholm University, the University of Maryland, Deutsches Elektronen-Synchrotron and Humboldt University, the TANGO Consortium of Taiwan, the University of Wisconsin at Milwaukee, Trinity College Dublin, Lawrence Livermore National Laboratories, IN2P3, University of Warwick, Ruhr University Bochum, Northwestern University, and former partners the University of Washington, Los Alamos National Laboratories, and Lawrence Berkeley National Laboratories. Operations are conducted by COO, IPAC, and UW.
ZTF access was supported by Northwestern University and the Center for Interdisciplinary Exploration and Research in Astrophysics (CIERA).
The SED Machine on P60 is based upon work supported by NSF grant 1106171.
Some of the data presented herein were obtained at Keck Observatory, which is a private 501(c)3 nonprofit organisation operated as a scientific partnership among the California Institute of Technology, the University of California, and the National Aeronautics and Space Administration. The Observatory was made possible by the generous financial support of the W. M. Keck Foundation. 
Based in part on observations collected at the European Organisation for Astronomical Research in the Southern Hemisphere under ESO programme(s) 105.20KC and 105.20PN. 
Data presented here were obtained in part with ALFOSC, which is provided by the Instituto de Astrof\'isica de Andaluc\'a (IAA) under a joint agreement with the University of Copenhagen and NOT. 
KAIT and its ongoing operation at Lick Observatory were made possible by donations from Sun Microsystems, Inc., the Hewlett-Packard Company, AutoScope Corporation, Lick Observatory, the U.S. NSF, the University of California, the Sylvia \& Jim Katzman Foundation, and the TABASGO Foundation. 
A major upgrade of the Kast spectrograph on the Shane 3\,m telescope at Lick Observatory was made possible through generous gifts from William and Marina Kast as well as the Heising-Simons Foundation. 
Research at Lick Observatory is partially supported by a generous gift from Google.   
Based on observations made with the Liverpool Telescope operated on the island of La Palma by Liverpool John Moores University in the Spanish Observatorio del Roque de los Muchachos of the Instituto de Astrofisica de Canarias with financial support from the UK Science and Technology Facilities Council.
We acknowledge the use of public data from the \swift\ data archive.

%% file: contributions.tex
\section{Author Contributions}

Contributors are sorted alphabetically.

\begin{itemize}

\item \textbf{Observations and Data Reduction} --- 
M. Bulla, R. Lunnan, S. Schulze (X-shooter), T.~G. Brink, A.~V. Filippenko, Y. Yang and W. Zheng (Keck, KAIT), K. Hinds and D.~A. Perley (LT), S. Schulze and J. Sollerman (NOT), Y. Sharma and T. Sit (P200), R. Lunnan, D.~A. Perley, Y. Sharma, and Y. Yao (Keck), S. Schulze ({\swift}), A. Gangopadhyay (bolometric light curve of SN 2020al), and K. Hinds and D.~A. Perley (BTS catalogue)

\item \textbf{Discoverer of the Si, S, Ar lines} --- 
A. Gal-Yam

\item \textbf{Analysis} --- 
P. Chen, L. Dessart, A. Gal-Yam, I. Irani, N. Kne\v zevi\'c, A.~A. Miller, D. A. Perley, N. Sarin, S. Schulze, N.~L. Strothjohannm D. Tsuna, and O. Yaron

\item \textbf{Discussion and Interpretation} --- 
All authors contributed to discussions and interpretation.

\item \textbf{Paper Writing} --- 
L. Dessart, A.~V. Filippenko, A. Gal-Yam, A.~A. Miller, S. Schulze, J. Sollerman, N.~L. Strothjohann, D. Tsuna, and S.~E. Woosley

\item \textbf{ZTF Infant SN Programme 2018--2023} --- R. J. Bruch, M. Bulla, P. Chen, S. Dhawan, A. Gal-Yam, I. Irani, S. Schulze, J. Sollerman, N.~L. Strothjohann, Y. Yang, O. Yaron, and E.~A. Zimmerman

\end{itemize}

%% file: supplementary_material.tex
\setcounter{figure}{0}
\setcounter{table}{0}

\section{Supplementary Material}
\section{Observations and Data Reduction}\label{app:obs}

\subsection{Photometry}\label{app:obs:photometry}

\indent\indent\textit{\textbf{Zwicky Transient Facility}} --- 
The Zwicky Transient Facility (ZTF) uses the Samuel Oschin 48-inch (1.22\,m) Schmidt telescope at Palomar Observatory on Mount Palomar (USA). It is equipped with a 47-square-degree camera \cite{Dekany2020a} and monitors the entire northern hemisphere every 2--3 days in the $g$ and $r$ bands to a depth of $\sim20.7$~mag ($5\sigma$; \cite{Bellm2019a, Graham2019a}) as  part of the public ZTF Northern Sky Survey \cite{Bellm2019b}. We retrieved the host-subtracted photometry via the Infrared Processing and Analysis Center (IPAC) ZTF forced-photometry service \cite{Masci2023a}. This service uses the data-reduction techniques outlined in Ref. \cite{Masci2019a}. We cleaned and calibrated the data following Ref. \cite{Masci2023a}.

\textit{\textbf{2.56\,m Nordic Optical Telescope}} --- 
We obtained photometry in $gri$ with the Alhambra Faint Object Spectrograph and Camera (ALFOSC)\footnote{\href{http://www.not.iac.es/instruments/alfosc}{{http://www.not.iac.es/instruments/alfosc}}} on the 2.56\,m Nordic Optical Telescope (NOT) at the Roque de los Muchachos Observatory on La Palma (Spain). To remove the host contribution, we obtained a final set of $gri$ photometry in August/September 2022, after the SN had faded. We reduced the data with \program{PyNOT}\footnote{\href{https://github.com/jkrogager/PyNOT}{https://github.com/jkrogager/PyNOT}} using standard techniques for CCD data processing and photometry. The world coordinate system was calibrated with the software package \program{astrometry.net} \cite{Lang2010a}. The host contribution was removed with  custom image-subtraction and analysis software (K. Hinds, K. Taggart, et al., in prep.). The photometry was measured using point-spread-function (PSF) fitting techniques based on methods in Ref. \cite{Fremling2016a}. 

\textit{\textbf{Palomar 60-inch telescope}} --- 
We acquired additional $ugri$ photometry using the Rainbow Camera of the Spectral Energy Distribution Machine (SEDM; \cite{Blagorodnova2018a, Rigault2019a}) on the robotic Palomar 60-inch (1.52\,m) telescope (P60; \cite{Cenko2006a}) at Palomar Observatory. The data were reduced using the data-reduction pipeline \program{FPipe} \cite[][]{Fremling2016a}.

\textit{\textbf{0.76\,m Katzman Automatic Imaging Telescope and 1\,m Nickel Telescope}} --- 
We obtained photometry in $BVRI$ and in the $Clear$ band (close to the $R$ band; \cite{Li2003a}) with the 0.76\,m Katzman Automatic Imaging Telescope (KAIT) at Lick Observatory on Mount Hamilton (USA) as a part of the Lick Observatory Supernova Search (LOSS; \cite{Filippenko2001a}). One additional epoch of photometry was also obtained with the 1\,m Nickel telescope at Lick Observatory. We reduced images using a custom pipeline\footnote{\href{https://github.com/benstahl92/LOSSPhotPypeline}{https://github.com/benstahl92/LOSSPhotPypeline}} detailed in Ref. \cite{Stahl2019a}. We performed PSF photometry with the package \program{DAOPHOT} \cite{Stetson1987a} from the \program{IDL Astronomy User's Library}\footnote{\href{http://idlastro.gsfc.nasa.gov}{http://idlastro.gsfc.nasa.gov}}.

\textit{\textbf{2\,m Liverpool Telescope}} --- 
We acquired photometry in $ugriz$ using the Infrared-Optical Imager (IO:O) on the robotic Liverpool Telescope (LT; \cite{Steele2004a}) at Roque de los Muchachos Observatory. Reduced images were downloaded from the LT archive and processed with custom image-subtraction and analysis software (K. Hinds, K. Taggart, et al., in prep.). Image stacking and alignment were performed using \program{SWarp} \cite{Bertin2002a} where required. Image subtraction was performed using a pre-explosion reference image in the appropriate filter from the Panoramic Survey Telescope and Rapid Response System (Pan-STARRS) Data Release (DR) 1 \cite{Chambers2016a} or Sloan Digital Sky Survey (SDSS) DR9 \cite{Ahn2012a}. The photometry is measured using PSF fitting techniques based on methods in Ref. \cite{Fremling2016a}. 

\textit{\textbf{Neil Gehrels Swift Observatory}} --- 
We submitted a target of opportunity request to use the 30\,cm Ultraviolet/Optical Telescope (UVOT; \cite{Roming2005a}) aboard the \textit{Neil Gehrels Swift Observatory} \cite{Gehrels2004a} to expand the wavelength coverage to the UV. Between January and February 2022, we obtained deep images in all filters to remove the host-galaxy contamination from the transient photometry. We coadded all sky exposures for a given epoch to boost the signal-to-noise ratio (S/N) using \program{uvotimsum} in \program{HEAsoft}\footnote{\href{https://heasarc.gsfc.nasa.gov/docs/software/heasoft}{https://heasarc.gsfc.nasa.gov/docs/software/heasoft}} version 6.32.2. Afterward, we measured the brightness of \sn with the \swift tool {\tt uvotsource}. The source aperture had a radius of $5''$, while the background region had a significantly larger radius. To remove the host contribution in $w2$, $m2$, and $w1$ from the earlier epochs, we arithmetically subtracted the host flux from the early measurements when the SN was bright.

\textit{\textbf{Final photometry}} --- The datasets were calibrated against stars from the Sloan Digital Sky Survey (SDSS; e.g., P60, LT, NOT observations; \cite{Ahn2012a}) and  Pan-STARRS (e.g., ZTF; \cite{Chambers2016a, Masci2019a, Flewelling2020a}), and internal zeropoints (\swift; \cite{Breeveld2011a}). Observations in similar but not identical filters (e.g., SDSS vs. ZTF filters) could introduce measurable, time-dependent colour terms \cite{Stritzinger2002a}. Convolving spectra between days 1.0 and 49.8 with SDSS and ZTF filter response functions yielded differences in the filter systems between 0.01 and 0.07 mag. They were comparable to, if not smaller than, the measurement uncertainties. Owing to this, we merged the datasets without applying any colour terms.

\begin{figure}
    \centering
    \captionsetup{name=\suppmattwo Fig.}
    \includegraphics[width=1\textwidth]{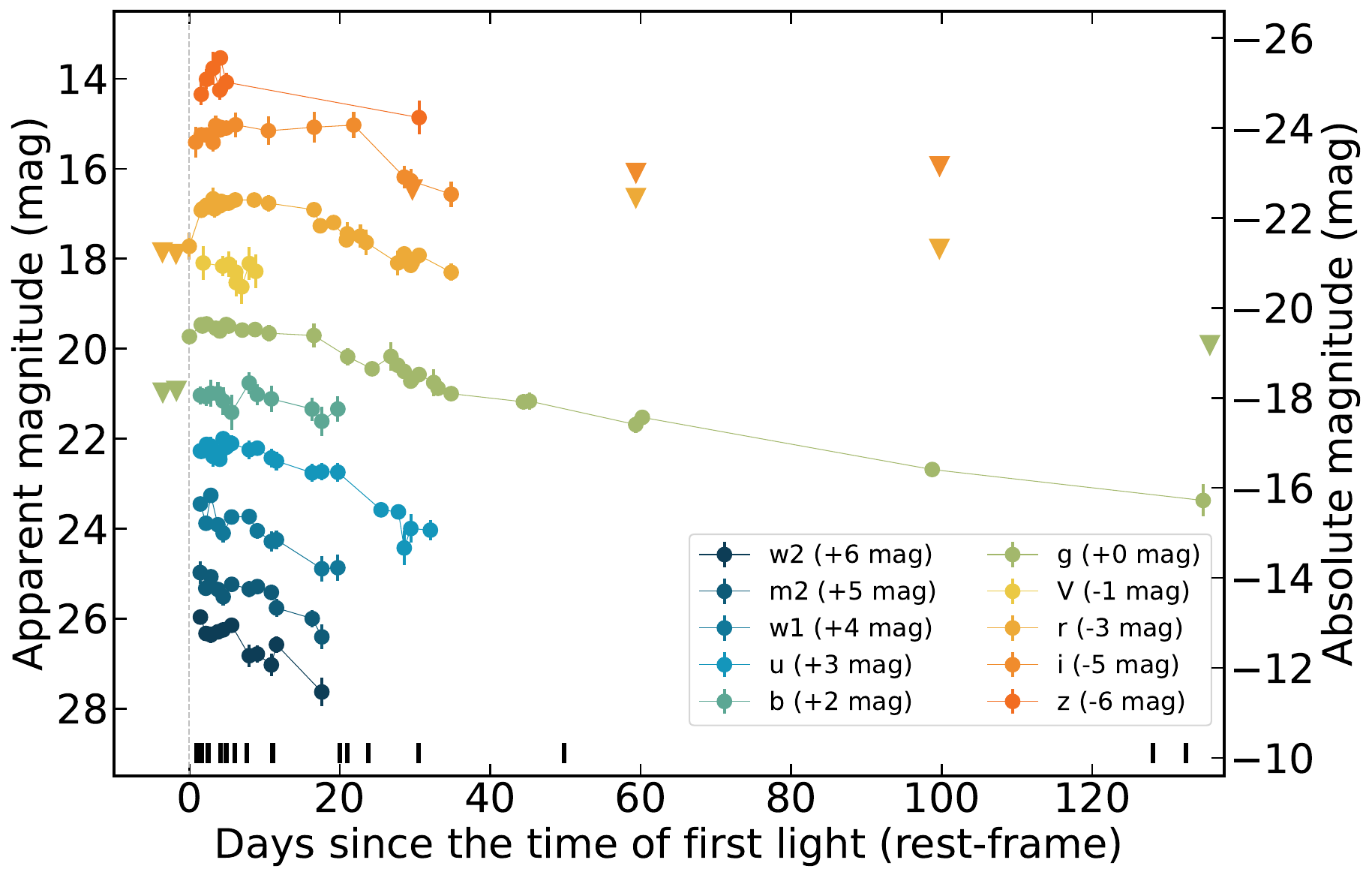}
    \caption{\textbf{The multiband light curves of \sn from 1,800 to 7,850~\AA\ (rest frame) corrected for MW extinction.} Vertical bars represent the epochs of spectroscopy. The absolute magnitude is computed with $M=m - {\rm DM}(z) + 2.5\,\log \left(1+z\right)$, where DM is the distance modulus and $z$ the redshift. Non-detections are displayed as `$\blacktriangledown$'.
    }
    \label{fig:lc}
\end{figure}

The final photometric data are shown in the \suppmat \figuretwo \ref{fig:lc} and the \suppmat \tabletwo \ref{tab:phot:sn}. All measurements are reported in the AB system \cite{Oke1983a}. The measurements in the \suppmat \tabletwo \ref{tab:phot:sn} are not corrected for Galactic extinction along the line of sight, but the Galactic extinction correction is applied to all photometric data shown in the figures and the derived properties. The MW extinction along the line of sight is $E(B-V)=0.03$ mag \cite{Schlafly2011a}. We assumed the Cardelli parameterisation of the MW extinction \cite{Cardelli1989a} and a total to selective extinction ratio of $R_V=3.1$.

\subsection{Spectroscopy}\label{app:obs:spectroscopy}

We obtained 16 spectra with several 2--10\,m-class telescopes. \suppmat \tabletwo \ref{tab:spec_log} shows the observing log. Details about the observations and the data reductions are provided below. 

\textit{\textbf{10\,m Keck Telescope}} ---
We obtained 4 epochs with the Low-Resolution Imaging Spectrometer (LRIS; \cite{Oke1995a}) on the 10\,m Keck I telescope at Maunakea (USA) between days 1.0 and  132.5. The first and the third epochs, acquired on 8 September 2021 (day 1) and 31 January 2022 (day 132.5), used the B600/4000 blue-side grism and the R400/8500 red-side grating,  dichroic 5600, and a $1\farcs0$-wide slit. For the second and fourth epochs (days 23.8 and 128.1), we utilised the B400/3400 blue-side grism and the R400/8500 red-side grating, dichroic 5600, and a $1\farcs0$-wide slit. The integration times varied between 1,200 and 3,600~s depending on the epoch. To minimise slit losses caused by atmospheric dispersion \citep{Filippenko1982a}, the spectra were acquired with the slit oriented at or near the parallactic angle. All spectra were reduced in a standard fashion with the data-reduction pipeline \program{LPipe} \cite{Perley2019a}.

\textit{\textbf{8.2\,m ESO Very Large Telescope}} --- 
We collected seven medium-resolution spectra with the X-shooter instrument \cite{Vernet2011a} at the 8.2\,m Very Large Telescope (VLT) at Paranal Observatory (Chile) between 9 September and 11 November 2021 (days 1.6 to 49.8). All observations were performed in nodding mode and with $1\farcs0$/$0\farcs9$/$0\farcs9$-wide slits (UVB/VIS/NIR arm). Each spectrum covers the wavelength interval from 3,000 to 24,800~\AA. The integration times varied between 2,880 and 4,400~s, depending on the arm (UVB/VIS/NIR) and phase. All observations were done with an atmospheric dispersion corrector to minimise any flux losses.
The data were reduced following Ref. \cite{Selsing2019a}. In brief, we first removed cosmic rays with the tool  \program{astroscrappy}\footnote{\href{https://github.com/astropy/astroscrappy}{https://github.com/astropy/astroscrappy}}, which is based on the cosmic-ray removal algorithm by Ref. \cite{vanDokkum2001a}. Afterward, the data were processed with the X-shooter pipeline v3.3.5 and the ESO workflow engine \program{ESOReflex} \cite{Goldoni2006a, Modigliani2010a}. The UVB- and VIS-arm data were reduced in stare mode to boost the S/N. The individual rectified, wavelength- and flux-calibrated 2D spectra files were coadded using tools developed by J. Selsing\footnote{\href{https://github.com/jselsing/XSGRB_reduction_scripts}{https://github.com/jselsing/XSGRB\_reduction\_scripts}}. The NIR data were reduced in nodding mode to ensure good sky-line subtraction. In the third step, we extracted the 1D spectra of each arm in a statistically optimal way using tools by J. Selsing. Finally, the wavelength calibration of all spectra was corrected for barycentric motion. The spectra of the individual arms were stitched by averaging the overlap regions.

\textit{\textbf{Palomar 200-inch Telescope}} ---
We obtained one epoch with the Double Spectrograph (DBSP; \cite{Oke1982a}) on the Palomar 200\,inch (5.1\,m) telescope at Mount Palomar Observatory on 13 September 2021 (day 5.2). The observations were taken using the D-55 dichroic beam splitter, a blue grating with 600 lines mm$^{-1}$ blazed at 4,000~\AA, a red grating with 316 lines mm$^{-1}$ blazed at 7,500~\AA, and a $1\farcs5$-wide slit. To minimise slit losses caused by atmospheric dispersion, the spectrum was acquired with the slit oriented at the parallactic angle. The data were reduced using the Python package \program{DBSP\_DRP}\footnote{\href{https://github.com/finagle29/dbsp_drp}{https://github.com/finagle29/dbsp\_drp}} that is primarily based on \program{PypeIt} \cite{Prochaska2020b, Prochaska2020a} and utilises common methods in optical spectroscopy.

\textit{\textbf{2.56\,m Nordic Optical Telescope}} ---
We collected 3 epochs of low-resolution spectroscopy with ALFOSC on the NOT between 7 January and 22 February 2019 (days 4.1 to 11.1). The spectra were obtained with grism \#4 and either a $1\farcs0$- or $1\farcs3$-wide slit, depending on the weather conditions. To minimise slit losses caused by atmospheric dispersion, the spectra were acquired with the slit oriented at the parallactic angle. The data were reduced with \program{PyNOT} using standard techniques for CCD data processing and long-slit spectroscopy. 

\textit{\textbf{Shane 3\,m telescope}} ---
We obtained one spectrum with the Kast double spectrograph\footnote{\href{https://mthamilton.ucolick.org/techdocs/instruments/kast/Tech\%20Report\%2066\%20KAST\%20Miller\%20Stone.pdf}{https://mthamilton.ucolick.org/techdocs/instruments/kast/Tech\%20Report\%2066\%20KAST\%20Miller\%20Stone.pdf}} mounted on the Shane 3\,m telescope at Lick Observatory on 11 September 2021 (day 3.5). We utilised a $2''$-wide slit, the 600/4310 grism in the blue, and the 300/7500 grating in the red. This instrument configuration has a combined wavelength range of $\sim3{,}500$--10,500~\AA.  To minimise slit losses caused by atmospheric dispersion, the Kast spectrum was acquired with the slit oriented at or near the parallactic angle. The Kast data were reduced following standard techniques for CCD processing and spectrum extraction \citep{Silverman2012a} utilising \program{IRAF} routines and custom \program{Python} and \program{IDL} codes\footnote{\href{https://github.com/ishivvers/TheKastShiv}{https://github.com/ishivvers/TheKastShiv}}. Owing to the low quality of the Lick spectrum, it is not shown in any of the figures in the paper, but it can be downloaded from WISeREP like all other spectra. 

\textbf{\textit{Flux calibration and host correction}} --- The flux calibration of all spectra was achieved with spectrophotometric standard stars observed during the same nights. We also tied the absolute flux scale to our multiband photometry. As the SN faded, the relative contribution from the host galaxy increased. To remove the host contribution, we used the X-shooter observation obtained at day 49.8, which also covered the host galaxy. The host was detected in the LRIS spectra from January/February 2022, too. While the continua of the three spectra were identical, the spectra differed in the relative amplitude of the emission lines due to the different resolving powers. 

\subsection{High-Energy Observations}\label{app:obs:xray}

While monitoring \sn with UVOT between day 1.5 and day 148.7, \swift also observed the field with its onboard X-ray telescope XRT between 0.3 and 10 keV in photon-counting mode \cite{Burrows2005a}. We analysed these data with the online tools of the UK \swift team\footnote{\href{https://www.swift.ac.uk/user_objects/}{https://www.swift.ac.uk/user\_objects}} that use the software package \program{HEASoft} version 6.26.1 and methods described in Refs. \cite{Evans2007a, Evans2009a}. \sn evaded detection at all epochs. The median $3\sigma$ count-rate limit of each observing block is $6\times10^{-3}~\rm s^{-1}$ (0.3--10~keV). Coadding all data pushes the $3\sigma$ count-rate limits to $0.6\times10^{-3}~\rm s^{-1}$. To convert the count-rate limits into a flux, we assume a power-law spectrum with a photon index\footnote{The photon index is defined as the power-law index of the photon flux density ($N(E)\propto E^{-\Gamma}$).} of $\Gamma=2$ and a Galactic neutral hydrogen column density of $2.7\times10^{20}$~cm$^{-2}$ \cite{HI4PI2016a}. The coadded count-rate limit translates to an unabsorbed flux of $<0.2\times10^{-13}~{\rm erg\,cm}^{-2}\,{\rm s}^{-1}$ in the range of 0.3--10 keV and a luminosity of $<1.2\times10^{42}~{\rm erg\,s}^{-1}$. \suppmat \tabletwo \ref{tab:xray} shows a list of all limits. 

Furthermore, we queried the NASA High Energy Astrophysics Science Archive Research Center (HEASARC\footnote{\href{https://heasarc.gsfc.nasa.gov/cgi-bin/W3Browse/w3browse.pl}{https://heasarc.gsfc.nasa.gov/cgi-bin/W3Browse/w3browse.pl}}) to search for any X-ray and $\gamma$-ray transient preceding or accompanying \sn. The source catalogues of the \textit{Fermi}, \textit{MAXI}, \textit{NICER}, \textit{NuSTAR}, \textit{AGILE}, and \textit{INTEGRAL} space missions returned no detection within $10'$ from \sn's position between 1 August 2021 (32.7~days before the discovery of \sn) and 4 February 2022 (131.8~days after the discovery of \sn). Placing a detection limit is not possible for this multitude of facilities.

\subsection{Host-galaxy Observations}\label{app:obs:host}

We retrieved science-ready coadded images from the \textit{Galaxy Evolution Explorer} (\galex) general release 6/7 \cite{Martin2005a}, the DESI Legacy Imaging Surveys (LS; \cite{Dey2018a}) DR 10, and PS1. We measured the brightness of the host using  \program{LAMBDAR} (Lambda Adaptive Multi-Band Deblending Algorithm in R; \cite{Wright2016a}) and the methods described in Ref. \cite{Schulze2021a}. The field was also observed by the VISTA Hemisphere Survey \cite{McMahon2013a} in the near-IR. We measured the brightness with the aperture photometry tool presented in Ref. \cite{Schulze2018a} using an aperture similar to the ones employed for the other images. The \galex, LS, and PanSTARRS photometry was calibrated against tabulated zeropoints, and the VHS photometry against stars from the 2MASS Point Source Catalogue \cite{Skrutskie2006a}. \suppmat \tabletwo \ref{tab:phot:host} summarises the measurements in the different bands.

The SN spectra also showed absorption lines from the ISM in the host and emission lines produced by the ionised gas in \ion{H}{ii}. \suppmat \tabletwo \ref{tab:host_ism} summarises the rest-frame equivalent widths of the absorption lines extracted from all early X-shooter spectra and the emission-line fluxes from X-shooter spectrum obtained at day 49.8. The flux measurements are not corrected for reddening.

\subsection{Comparison Objects}\label{app:literature}

We compare the spectra and light curves of \sn to those of other objects. Below, we list the relevant references for each object.

\begin{itemize}
    \item SN\,2006jc (Ibn) --- bolometric light curve, not constructed; spectra, Ref. \cite{Pastorello2007a}
    \item SN\,2010al (Ibn) --- bolometric light curve, built with data from Ref. \cite{Pastorello2015a} using the programme \program{Superbol} \cite{Nicholl2018b}; spectra, Ref. \cite{Pastorello2015a}
    \item SN\,2019hgp (Icn) --- bolometric light curve, Ref. \cite{Gal-Yam2022a}; spectra, Ref. \cite{Gal-Yam2022a}
    \item SN\,2020oi (Ic) --- bolometric light curve, Ref. \cite{Gagliano2022a}
    \item SN\,2021csp (Icn) --- bolometric light curve, Ref. \cite{Perley2022a}; spectra, Ref. \cite{Perley2022a}
\end{itemize}

\noindent All spectra were retrieved from WISeREP.

\clearpage

\clearpage
\input{tables/tab_lines}
\input{tables/SN2021yfj_photometry_final.tex}
\clearpage
\input{tables/tab_spec}
\clearpage
\input{tables/tab_xray.tex}

\clearpage
\input{tables/tab_host}
\input{tables/tab_host_ism.tex}
\clearpage
\input{tables/tab_lbol}
\clearpage
\input{tables/tab_redback_v2}
\input{tables/tab_chips}

%% file: tables/tab_lines.tex
\begin{table}
\captionsetup{name=\suppmattwo Table}
\caption{Lines identified in the spectra shown in \figuretwo \ref{fig:spec:flash:id} and the \extmat \figuretwo \ref{fig:spec:flash:id:full}.\label{tab:lines}}
\centering
\RaggedRight
\begin{tabular}{ c m{10cm}}
\toprule
Species & Lines (\AA) \\
\midrule
\vspace{1mm}
\ion{Ar}{iii} & 3285.8, 3301.9, 3311.2, 3336.2, 3344.8, 3480.5, 3511.2, 3514.2\\
\vspace{1mm}
\ion{C}{iii}  & 4647.4, 4650.3, 4651.5, 7771.8\\
\vspace{1mm}
\ion{He}{i}   & 3888.6, 5875.6, 7065.2, 10830 \\
\vspace{1mm}
\ion{Mg}{ii}  & 2795.5 , 2802.7 \\
\vspace{1mm}
\ion{S}{iii}  & 3324.0, 3324.8, 3367.2, 3369.5, 3370.4, 3387.1, 3497.3, 3499.2, 3632.0, 3661.9, 3710.4, 3717.7, 3747.9, 3750.7, 3794.6, 3837.7, 3838.3, 3860.6, 3928.6, 3983.7, 3985.9, 4091.2, 4253.5, 4284.9, 4354.5, 4361.5, 4364.7, 4418.8, 4677.6, 5160.1, 5219.3\\
\vspace{1mm}
\ion{S}{iv}   & 3097.3, 3117.6, 3119.9, 3338.6, 3340.4, 3341.5\\
\vspace{1mm}
\ion{Si}{iii} & 3096.8, 3185.1, 3241.6, 3486.8, 3590.5, 3796.1, 3806.5, 4552.6, 4567.8, 4574.8, 4716.7, 4813.3, 4819.7, 4828.95, 5739.73, 7461.9, 7466.3\\
\vspace{1mm}
\ion{Si}{iv}  & 3149.6, 3165.7, 4088.9, 4116.1\\
\toprule
\end{tabular}
{\raggedright \textbf{Notes:} These are the Ritz air wavelengths reported in the NIST atomic spectra database (details at \href{https://physics.nist.gov/PhysRefData/ASD/Html/lineshelp.html}{https://physics.nist.gov/PhysRefData/ASD/Html/lineshelp.html}). All wavelengths are rounded to one decimal place for convenience.\par}
\end{table}

%% file: tables/SN2021yfj_photometry_final.tex
\begin{table}
\captionsetup{name=\suppmattwo Table}
\caption{Log of photometric observations.}\label{tab:phot:sn}
\begin{tabular}{ccccc}
\toprule
MJD & Phase & Telescope/ & Filter & Brightness \\ 
    & (day) & Instrument &        & (mag)      \\ 
\midrule
59466.102    &$    1.5    $&   \swift/UVOT &$    w2    $&$    20.23    \pm    0.11    $\\
59466.951    &$    2.2    $&   \swift/UVOT &$    w2    $&$    20.59    \pm    0.14    $\\
59467.683    &$    2.9    $&   \swift/UVOT &$    w2    $&$    20.62    \pm    0.18    $\\
59468.744    &$    3.8    $&   \swift/UVOT &$    w2    $&$    20.56    \pm    0.16    $\\
59469.546    &$    4.5    $&   \swift/UVOT &$    w2    $&$    20.51    \pm    0.15    $\\
59470.830    &$    5.6    $&   \swift/UVOT &$    w2    $&$    20.41    \pm    0.15    $\\
\bottomrule
\end{tabular}
{\raggedright \textbf{Notes:} All measurements are reported in the AB system and are not corrected for reddening. The phase is reported in the rest-frame with respect to the time of the first detection (MJD=59464.414). Non-detections are reported at $3\sigma$ confidence. A machine-readable table will be made available via the WISeREP archive and the journal webpage after the acceptance of the paper.}
\end{table} 

%% file: tables/tab_spec.tex
\begin{sidewaystable}
\captionsetup{name=\suppmattwo Table}
\caption{Log of spectroscopic observations.\label{tab:spec_log}}
\tiny
\begin{tabular*}{\textheight}{@{\extracolsep\fill}lccccccc}
\toprule
MJD & Phase & Telescope/ & Disperser & Slit          & Exposure & Wavelength  & Spectral  \\
    & (day) & Instrument &           & width ($''$)  & time (s) & range (\AA) & resolution\\
\midrule
59,465.57 &$ 1.0   $& Keck-I/LRIS   & 600/4000, 400/8500   & 1\farcs0                   & 1,230             & 3,400 -- 10,275 & 1,000/1,200       \\
59,466.28 &$ 1.6   $& VLT/X-shooter & ...                  & 1\farcs0/0\farcs9/0\farcs9 & 3,000/2,900/2,880 & 3,000 -- 24,800 & 5,400/8,900/5,600 \\
59,467.28 &$ 2.5   $& VLT/X-shooter & ...                  & 1\farcs0/0\farcs9/0\farcs9 & 3,000/2,900/2,880 & 3,000 -- 24,800 & 5,400/8,900/5,600 \\
59,468.39 &$ 3.5   $& Lick/Kast     & 600/4310, 300/7500   & 2\farcs0, 2\farcs0         & 3,660/3,600       & 3,500 -- 10,500 & 800               \\
59,469.12 &$ 4.1   $& NOT/ALFOSC    & Gr\#4                & 1\farcs3                   & 4,500             & 3,600 -- 9,600  & 280               \\
59,470.00 &$ 5.2   $& P200/DBSP     & 600/4000, 316/7500   & 1\farcs5                   & 1,800             & 3,500 -- 10,000 & 1,000/1,000       \\
59,471.31 &$ 6.1   $& VLT/X-shooter & ...                  & 1\farcs0/0\farcs9/0\farcs9 & 3,000/2,900/2,880 & 3,000 -- 24,800 & 5,400/8,900/5,600 \\
59,473.16 &$ 7.7   $& NOT/ALFOSC    & Gr\#4                & 1\farcs0                   & 4,500             & 3,600 -- 9,600  & 360               \\
59,477.05 &$ 11.1  $& NOT/ALFOSC    & Gr\#4                & 1\farcs3                   & 4,500             & 3,600 -- 9,600  & 280               \\
59,487.21 &$ 20.0  $& VLT/X-shooter & ...                  & 1\farcs0/0\farcs9/0\farcs9 & 3,600/3,716/3,600 & 3,000 -- 24,800 & 5,400/8,900/5,600 \\
59,488.30 &$ 21.0  $& VLT/X-shooter & ...                  & 1\farcs0/0\farcs9/0\farcs9 & 3,000/2,900/2,880 & 3,000 -- 24,800 & 5,400/8,900/5,600 \\
59,491.52 &$ 23.8  $& Keck-I/LRIS   & 400/3400, 400/8500   & 1\farcs0                   & 1,800             & 3,076 -- 9,350  & 600/1,200         \\
59,499.10 &$ 30.5  $& VLT/X-shooter & ...                  & 1\farcs0/0\farcs9/0\farcs9 & 4,400/4,400/3,840 & 3,000 -- 24,800 & 5,400/8,900/5,600 \\
59,521.13 &$ 49.8  $& VLT/X-shooter & ...                  & 1\farcs0/0\farcs9/0\farcs9 & 4,400/4,400/3,840 & 3,000 -- 24,800 & 5,400/8,900/5,600 \\
59,610.25 &$ 128.1 $& Keck-I/LRIS   & 600/4000, 400/8500   & 1\farcs0                   & 3,600             & 3,400 -- 10,275 & 1,000/1,200       \\
59,615.27 &$ 132.5 $& Keck-I/LRIS   & 400/3400, 400/8500   & 1\farcs0                   & 2,700             & 3,076 -- 9,350  & 600/1,200         \\
\botrule
\end{tabular*}
{\raggedright \textbf{Notes:} 
The modified Julian dates quote the beginning of each observation. The phase is reported for the mid-exposure time and in the rest frame with respect to the time of the first detection (MJD = 59,464.414). For the multi-arm instruments Kast, LRIS, and X-shooter, we report the exposure time and the spectral resolution of each arm. The wavelength ranges and the values of the spectral resolutions were taken from instrument manuals. \par
}
\end{sidewaystable}

%% file: tables/tab_xray.tex
\begin{table}
\captionsetup{name=\suppmattwo Table}
\caption{Log of X-ray observations.\label{tab:xray}}
\centering
\begin{tabular}{cccccc}
\toprule
MJD & Phase & Count rate               & $F_{\rm X}$                             & $L_{\rm X}$        \\
    & (day) & ($10^{-3}~\rm s^{-1}$)   & ($10^{-13}~\rm erg\,s^{-1}\,cm^{-2}$)   & ($10^{42}~\rm erg\,s^{-1}$)\\
\midrule
\multicolumn{5}{c}{\textbf{Unbinned data}}\\
\midrule
$59,466.10 \pm 0.04 $&$ 1.5    $&$<18.6	$&$<6.95	$&$<38.01$\\ 
$59,466.95 \pm 0.04 $&$ 2.2    $&$<6.0	$&$<2.24	$&$<12.24$\\ 
$59,467.68 \pm 0.30 $&$ 2.9    $&$<11.4	$&$<4.28	$&$<23.42$\\ 
$59,468.74 \pm 0.10 $&$ 3.8    $&$<5.7	$&$<2.14	$&$<11.68$\\ 
$59,469.55 \pm 0.04 $&$ 4.5    $&$<32.4	$&$<12.15	$&$<66.41$\\ 
$59,470.83 \pm 0.01 $&$ 5.6    $&$<7.4	$&$<2.78	$&$<15.21$\\ 
$59,473.49 \pm 0.07 $&$ 8.0    $&$<11.7	$&$<4.37	$&$<23.90$\\ 
$59,474.71 \pm 0.04 $&$ 9.0    $&$<4.7	$&$<1.75	$&$<9.59 $\\
$59,476.84 \pm 0.04 $&$ 10.9   $&$<5.5	$&$<2.06	$&$<11.25$\\ 
$59,477.61 \pm 0.07 $&$ 11.6   $&$<4.7	$&$<1.76	$&$<9.61 $\\
$59,483.02 \pm 0.04 $&$ 16.3   $&$<4.1	$&$<1.55	$&$<8.47 $\\
$59,484.48 \pm 0.44 $&$ 17.6   $&$<3.9	$&$<1.46	$&$<7.97 $\\
$59,486.90 \pm 0.07 $&$ 19.7   $&$<4.6	$&$<1.73	$&$<9.44 $\\
$59,493.44 \pm 0.23 $&$ 25.5   $&$<11.6	$&$<4.34	$&$<23.72$\\ 
$59,496.06 \pm 0.01 $&$ 27.8   $&$<11.7	$&$<4.39	$&$<23.98$\\ 
$59,500.90 \pm 0.01 $&$ 32.0   $&$<9.4	$&$<3.51	$&$<19.18$\\ 
$59,581.30 \pm 0.23 $&$ 102.7  $&$<2.2	$&$<0.83	$&$<4.55 $\\
$59,584.52 \pm 0.01 $&$ 105.5  $&$<17.7	$&$<6.65	$&$<36.33$\\ 
$59,633.42 \pm 0.20 $&$ 148.4  $&$<4.7	$&$<1.78	$&$<9.72 $\\
\midrule
\multicolumn{5}{c}{\textbf{Binned data}}\\
\midrule
$59,475.18^{+158.45}_{-9.12}$&$ 9.5^{+139.2}_{-8.0}$&$<0.6$&$<0.22$&$<1.22$\\
\bottomrule
\end{tabular}
{\raggedright \textbf{Notes:} The modified Julian dates report the mid-exposure time. The phase is reported in the rest-frame with respect to the time of the first detection (MJD = 59,464.414). The time errors indicate the extent of the time bins. All limits are reported at $3\sigma$ confidence. The measurements are corrected for MW absorption and are reported for the bandpass from 0.3 to 10~keV. \par
}
\end{table}

%% file: tables/tab_host.tex
\begin{table}[h]
\captionsetup{name=\suppmattwo Table}
\caption{Photometry of the host galaxy.\label{tab:phot:host}}
\begin{tabular}{ccclcc}
\toprule
Survey & Filter & Brightness & Survey & Filter & Brightness\\
       &        & (mag)      &        &        & (mag)     \\
\midrule
\galex        & $FUV$       &$ 23.04 \pm 0.19$ & PS1     & $g$         &$ 21.58 \pm 0.11$\\
\galex        & $NUV$       &$ 22.94 \pm 0.18$ & PS1     & $r$         &$ 20.99 \pm 0.15$\\
LS & $g$         &$ 21.45 \pm 0.02$ & PS1     & $i$         &$ 20.73 \pm 0.08$\\
LS & $r$         &$ 20.97 \pm 0.02$ & VHS           & $J         $&$ 20.85 \pm 0.12$\\
LS & $i$         &$ 20.68 \pm 0.03$ & VHS           & $H         $&$ 20.46 \pm 0.09$\\
LS & $z$         &$ 20.59 \pm 0.03$ & VHS           & $K_{\rm s} $&$ 21.22 \pm 0.33$\\
\botrule
\end{tabular}
{\raggedright \textbf{Notes:} 
All measurements are reported in the AB system and are not corrected for reddening. \par
}
\end{table}

%% file: tables/tab_host_ism.tex
\begin{table}
\captionsetup{name=\suppmattwo Table}
\caption{Properties of the ISM in the host galaxy.}\label{tab:host_ism}
\centering
\begin{tabular}{lcc}
\toprule
Transition                 & ${\rm EW}_{\rm r}$  & Flux \\
                           & (\AA)                                         & $\left(10^{-16}\,{\rm erg\,cm}^{-2}\,{\rm s}^{-1}\right)$\\
\midrule
\multicolumn{3}{c}{\textbf{Absorption lines}}\\
\midrule
\ion{Mg}{ii}\,$\lambda$\,2804    & $1.02  \pm 0.06$  & \nodata     \\
\ion{Mg}{i}\,$\lambda$\,2852     & $0.30  \pm 0.08$  & \nodata     \\
\midrule
\multicolumn{3}{c}{\textbf{Emission lines}}\\
\midrule
{[\ion{O}{ii}]}\,$\lambda$\,3926    & \nodata       & $ 1.62	\pm 0.04$ \\
{[\ion{O}{ii}]}\,$\lambda$\,3929    & \nodata       & $ 2.36   \pm 0.04$ \\
{[\ion{Ne}{iii}]}\,$\lambda$\,3869   & \nodata       & $ 0.25   \pm 0.02$ \\
H$\epsilon$                         & \nodata       & $ 0.12   \pm 0.02$ \\
H$\delta$                           & \nodata       & $ 0.27   \pm 0.02$ \\
H$\gamma$                           & \nodata       & $ 0.55   \pm 0.02$ \\
{[\ion{O}{iii}]}\,$\lambda$\,4363$^\dagger$   & \nodata       & $ 0.04   \pm 0.03$ \\
H$\beta$                            & \nodata       & $ 1.21   \pm 0.03$ \\
{[\ion{O}{iii}]}\,$\lambda$\,4959   & \nodata       & $ 0.84   \pm 0.04$ \\
{[\ion{O}{iii}]}\,$\lambda$\,5007   & \nodata       & $ 2.57   \pm 0.06$ \\
{[N\textsc{ii}]}\,$\lambda$\,6549   & \nodata       & $ 0.23   \pm 0.02$ \\
H$\alpha$                           & \nodata       & $ 4.87   \pm 0.05$ \\
{[N\textsc{ii}]}\,$\lambda$\,6584   & \nodata       & $ 0.66   \pm 0.02$ \\
{[\ion{S}{ii}]}\,$\lambda$\,6718    & \nodata       & $ 0.92   \pm 0.03$ \\
{[\ion{S}{ii}]}\,$\lambda$\,6732    & \nodata       & $ 0.62   \pm 0.03$ \\
\bottomrule
\end{tabular}
{\raggedright \textbf{Notes:} We report rest-frame equivalent widths EW$_{\rm r}$ for absorption lines and fluxes for emission lines.  The emission lines are measured from the X-shooter spectrum at day 49.8 and are not corrected for reddening.\\
$^\dagger$ Measurement has a significance of $<3\sigma$.
}
\end{table}

%% file: tables/tab_lbol.tex
\begin{table}
\captionsetup{name=\suppmattwo Table}
\caption{The bolometric light curve and blackbody properties. }\label{tab:bolometrics}
\begin{tabular}{cccccc} 
\toprule
Phase & $\log~L_{\rm bol}             $& $\log~L_{\rm bol}             $& $T_{\rm BB}$ & $\log~R_{\rm BB}$& Bolometric  \\
(day) & $\left(\rm erg\,s^{-1}\right) $& $\left(\rm erg\,s^{-1}\right) $& (K)          & (cm)             & correction? \\
      & (1,800--7,850~\AA)               & (+FUV +IR)                     &              &                  &             \\
\midrule
$0    $&$ 43.38\pm0.09 $&$ 43.60\pm0.08 $&$ \nodata        $&$ \nodata               $& yes  \\
$0.28 $&$ 43.42\pm0.07 $&$ 43.62\pm0.07 $&$ \nodata        $&$ \nodata               $& yes  \\
$0.78 $&$ 43.45\pm0.06 $&$ 43.64\pm0.06 $&$ \nodata        $&$ \nodata               $& yes  \\
$1.28 $&$ 43.49\pm0.05 $&$ 43.67\pm0.05 $&$ \nodata        $&$ \nodata               $& yes  \\
$1.78 $&$ 43.52\pm0.04 $&$ 43.69\pm0.05 $&$ 15,522 \pm 998 $&$ 15.00 \pm 0.04        $& no   \\
$2.28 $&$ 43.52\pm0.04 $&$ 43.68\pm0.05 $&$ 15,351 \pm 936 $&$ 15.00 \pm 0.04        $& no   \\
\bottomrule
\end{tabular}
{\raggedright \textbf{Notes:} The phase is reported in the rest frame with respect to the time of the first detection (MJD = 59,464.414). The last column indicates whether a bolometric correction was applied to determine the pseudobolometric luminosity in the wavelength interval 1,800--7,850~\AA. Including the missing FUV and IR flux required a bolometric correction for all data (see the \method \nameref{app:lc:bolometrics} for details). The blackbody properties are only reported where UV and optical photometry are available and where the spectrum is adequately described by a blackbody. A machine-readable table will be made available via the WISeREP archive and the journal webpage after the acceptance of the paper.}
\end{table} 

%% file: tables/tab_redback_v2.tex
\begin{table}[h]
\captionsetup{name=\suppmattwo Table}
\caption{Light-curve fits with \program{Redback}: models, priors and marginalised posteriors.\label{tab:lc:redback}}

\begin{tabular}{lcrr}
\toprule
Parameter & Prior &  Magnetar  & Nickel\\
\midrule
\multicolumn{4}{c}{\textbf{General}} \\
\midrule
ejecta mass $M_{\rm ej}$ $\left(M_\odot\right)$                         & $\log \mathcal{U}\left(0.01, 10\right)$             &$0.3^{+0.3}_{-0.1}     $&$1.7\pm0.1            $\\
explosion date $t_{\rm exp}$ (day, observer frame)                      & $\mathcal{U}\left(-10, 0\right)$                    &$-4.5^{+1.0}_{-0.3}    $&$-8.3\pm0.5           $\\
``$\gamma$-ray'' opacity $\kappa_\gamma$ $\left(\rm cm^2\,g^{-1}\right)$& $\log \mathcal{U}\left(10^{-4}, 10^{4}\right)      $&$0.09^{+0.06}_{-0.03}  $&$0.034$ (fixed)        \\
optical opacity $\kappa$ $\left(\rm cm^2\,g^{-1}\right)$                & $\mathcal{U}\left(0.01, 0.5\right)$                 &$0.01 \pm0.01          $&$0.07$ (fixed)         \\
photospheric plateau temperature $T$~(K)                                & $\mathcal{U}\left(5000, 15{,}000\right)$            &$14{,}600^{+300}_{-400}$&$14{,}200\pm300       $\\
white-noise parameter $\sigma$                                          & $\mathcal{U}\left(10^{-3}, 2\right)$                &$0.01                  $&$0.01                 $\\
$V$-band absorption $A_{\rm V}$ (mag)                                   & $\mathcal{U}\left(0, 1\right)$                      &$0.08^{+0.08}_{-0.05}  $&$0.01\pm0.01$          \\
\midrule
\multicolumn{4}{c}{{\textbf{Magnetar model}}} \\
\midrule
ejecta nickel mass fraction $f_{\rm Ni}$                                & $\log \mathcal{U}\left(10^{-3}, 1\right)$           &$0.05^{+0.32}_{-0.05}        $&\dots\\
initial spin-down luminosity $L_0$ ($10^{44}~\rm erg\,s^{-1}$)          & $\log \mathcal{U}\left(10^{-4}, 10^{6}\right)$      &$0.9^{+2.1}_{-0.2}           $&\dots\\
braking index $n$                                                       & $\mathcal{U}\left(1.5, 10\right)$                   &$7.9^{+1.4}_{-1.8}           $&\dots\\
spin-down time $\tau_{\rm sd}$ (s)                                      & $\log \mathcal{U}\left(10, 10^8\right)$             &$1.3^{+0.8}_{-1.0}\times10^6 $&\dots\\
supernova explosion energy $E_{\rm SN}$ ($10^{50}~\rm erg$)             & $\mathcal{U}\left(1, 50\right)$                     &$1.9^{+1.0}_{-0.6}           $&\dots\\
\midrule
\multicolumn{4}{c}{\textbf{$^{56}$Ni model}} \\
\midrule
ejecta nickel mass fraction $f_{\rm Ni}$                                & $\log \mathcal{U}\left(10^{-3}, 1\right)$           &\dots                   &$0.97\pm0.01           $\\
scaling velocity $v_{\rm scale}$ $\left({\rm km\,s}^{-1}\right)$        & $\mathcal{U}\left(1000, 15{,}000\right)$            &\dots                   &$14{,}300^{+400}_{-600}$\\
\midrule
\multicolumn{4}{c}{\textbf{Fit quality}}\\
\midrule
log Bayesian evidence ($\log~Z$)                                        &                                                     & 547                  & 518                          \\
number of free parameters                                               &                                                     &  12                  &   7                          \\
\bottomrule
\end{tabular}
{\raggedright \textbf{Notes:} We used uniform ($\mathcal{U}$) and log-uniform ($\log \mathcal{U}$) priors. The uncertainties of the marginalised posteriors are quoted at $1\sigma$ confidence. The explosion date is measured with respect to the time of the first detection. All marginalised posteriors are reported in linear units. The Bayesian evidence is reported in log units. The opacities of the nickel model were taken from Refs. \cite{Swartz1995a, Wang2015a}.
}
\end{table}

%% file: tables/tab_chips.tex
\begin{table}
\captionsetup{name=\suppmattwo Table}
\caption{Light curve fit with \program{CHIPS}.}\label{tab:chips}
\begin{tabular}{lc}
\toprule
Parameter & Value \\
\midrule
ejecta mass $M_{\rm ej}$ $\left(M_\odot\right)$                   & 5                  \\
CSM mass    $M_{\rm CSM}$ $\left(M_\odot\right)$                  & $>1$               \\
explosion energy $E_{\rm ej}$ (erg)                               & $1.6\times 10^{51}$\\
power-law index of the inner CSM $n_{\rm in}$                     & $1.1$              \\
radius at transition $r_{\rm CSM}$ (cm)                           & $5\times 10^{15}$  \\
density at transition $\hat{\rho}_{\rm CSM}$ $({\rm g\ cm}^{-3})$ & $3\times 10^{-15}$ \\
\bottomrule
\end{tabular}
\footnotetext{}
\end{table}